\documentclass[manuscript,screen]{acmart}

\usepackage{graphicx}
\usepackage{tikz}

\usepackage{caption}
\usepackage{subcaption}
\usepackage{tabu}
\usepackage{multirow}
\usepackage{makecell}

\usepackage{xcolor, soul}

\AtBeginDocument{%
  \providecommand\BibTeX{{%
    \normalfont B\kern-0.5em{\scshape i\kern-0.25em b}\kern-0.8em\TeX}}}

\setcopyright{acmcopyright}
\copyrightyear{2018}
\acmYear{2018}
\acmDOI{XXXXXXX.XXXXXXX}

\acmJournal{JACM}
\acmVolume{37}
\acmNumber{4}
\acmArticle{111}
\acmMonth{8}

\begin{document}




\title[A Survey on Deep Learning for Design and Generation of Virtual Architecture]
{Towards AI-Architecture Liberty: A Comprehensive Survey on Design and Generation of Virtual Architecture by Deep Learning}



\author{Anqi Wang}
\affiliation{\institution{Emerging Interdisciplinary Areas, Hong Kong University of Science and Technology}\country{Hong Kong SAR}}
\affiliation{\institution{Computational Media and Arts, Hong Kong University of Science and Technology (Guangzhou)}\city{Guangzhou}\country{China}}

\author{Jiahua Dong}
\affiliation{\institution{The Chinese University of Hong Kong}\country{Hong Kong SAR}}

\author{Lik-Hang Lee}
\affiliation{\institution{The Hong Kong Polytechnic University}\country{Hong Kong SAR}}

\author{Jiachuan Shen}
\affiliation{\institution{The Bartlett School of Architecture, University College London}\city{London}\country{UK}}

\author{Pan Hui}
\affiliation{\institution{Computational Media and Arts, Hong Kong University of Science and Technology (Guangzhou)}\city{Guangzhou}\country{China}}
\affiliation{\institution{Emerging Interdisciplinary Areas, Hong Kong University of Science and Technology}\country{Hong Kong SAR}}


\begin{abstract}

3D shape generation techniques leveraging deep learning have garnered significant interest from both the computer vision and architectural design communities, promising to enrich the content in the virtual environment. However, research on virtual architectural design remains limited, particularly regarding designer-AI collaboration and deep learning-assisted design.
In our survey, we reviewed 149 related articles (81.2\% of articles published between 2019 and 2023) covering architectural design, 3D shape techniques, and virtual environments.
Through scrutinizing the literature, we first identify the principles of virtual architecture and illuminate its current production challenges, including datasets, multimodality, design intuition, and generative frameworks.
We then introduce the latest approaches to designing and generating virtual buildings leveraging 3D shape generation and summarize four characteristics of various approaches to virtual architecture.
Based on our analysis, we expound on four research agendas, including agency, communication, user consideration, and integrating tools.
Additionally, we highlight four important enablers of ubiquitous interaction with immersive systems in deep learning-assisted architectural generation. 
Our work contributes to fostering understanding between designers and deep learning techniques, broadening access to designer-AI collaboration. We advocate for interdisciplinary efforts to address this timely research topic, facilitating content designing and generation in the virtual environment.
\end{abstract}



\maketitle
\section{Introduction}

In recent decades, the exploration of architectural space has shifted from reinforced concrete to digital architecture and further to information frameworks, with virtualization extending beyond the physical layer~\cite{carpo2017second,bava_2020}. Architecture is no longer perceived as static, permanent objects but as integral components of a data network and evolving communication between different architectural systems~\cite{claypool_2019, SCHROEDER2001569}.
Digital innovations and technological advancements associated with spline, pixels, voxels, and bits have redefined architectural forms called digital architecture~\cite{jovanovic_2022}. 
In such context, two powerful forces have driven the birth of generated virtual architecture: AI-assisted design and virtual reality (VR) technique. 
In recent decades, AI architecture has become a critical sub-genre of architectural discourse advocated by scholars and practitioners~\cite{stojanovski_rethinking_2022},
utilizing deep learning neural networks to concern social issues. Moreover, the burgeoning innovates and techniques of virtual environments (e.g. virtual scenes with virtual reality (VR), augmented reality (AR), immersive videogames, and social VR) have expanded the boundaries of production and design in digital fabrication buildings, such as physical-virtual merging~\cite{cheng2019VRoamer,lee2021towards,lin_expanding_2021}, bio-signal feedback~\cite{ding_spatial_2022,pei_biofeedback_2020,homolja2020impact,pei_detecting_2021}, and participatory design in the virtual environment~\cite{kim_placemakingai_2022,kwiecinski_participatory_2017}.
Thus, deep learning facilitates the design and generation of diverse virtual architectures with interactive features, supporting the promising vision of ubiquitous intelligent environments in VR. It simulates real-world scenarios by generating and utilizing 3D building datasets, aligning design objectives with human needs~\cite{doi:10.1177/14780771221100102}.


\begin{figure}[t]
    \centering
    \includegraphics[width=.98\textwidth]{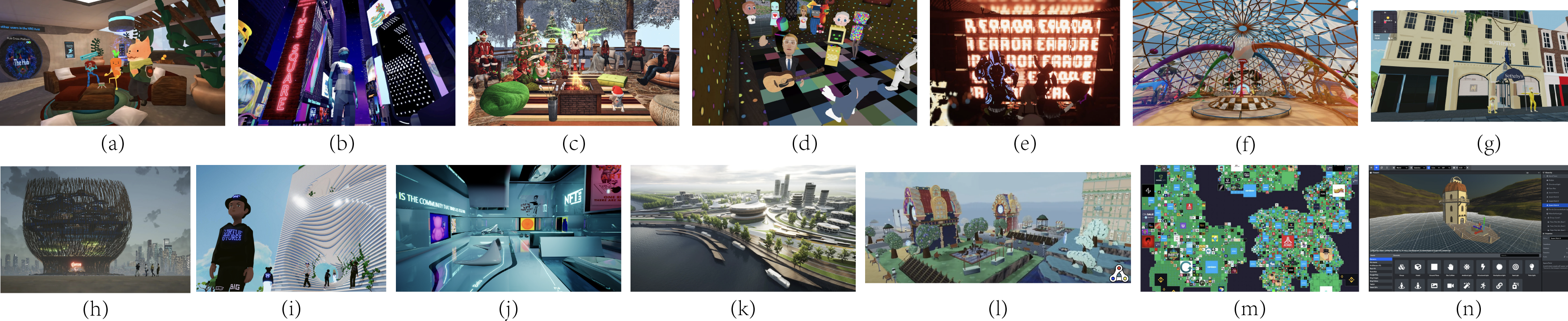}     

    \caption
    {Virtual architecture provides rich use affordances, benefiting various aspects, including users, communities, and organizations.
        (a) Hub Rooms in VRChat;
        (b) Simulated city scene with various buildings;
        (c) Interior of individual house in a virtual world game, Second Life;
        (d) Saint Motel Fan Meet on Mozzila hub; 
        (e) Sci-fi nightclub;  
        (f) Center building in pride month event held on Dencentraland; 
        (g) Realism Sotheby’s virtual gallery;
        (h) The interactive digital experience in a virtual building for the brand Jose Cuervo designed by Rojkind Arquitectos;
        (i) The office building of Vice Media Group was designed in metaverse by BIG Architect;
        (j) "NFTism" is a virtual NFT gallery developed collaboratively by Zaha Hadid Architects (ZHA) and Journee. The architectural design focuses on user experience, social interaction, and "dramaturgical" compositions, supporting MMO (massively multiplayer online) technology and integrating audio-video interaction\protect\footnotemark; 
        (k) The virtual city designed and developed by ZHA, filled with fluid style virtual architecture; 
        (l) Each piece of land and each item in the virtual land as a non-fungible token could trade in the virtual world of Decentraland; 
        (m) A mapping showing ownership of non-fungible token land in Sandbox;
        (n) A user-friendly tool for editing and creating buildings and land in social VR, Mozzila hub.}
        \label{fig:VAusecase}
\end{figure}
\footnotetext{A report by Archdaily, source:~\protect\url{https://www.archdaily.com/972886/zaha-hadid-architects-presents-virtual-gallery-exploring-architecture-nfts-and-the-metaverse}}

The designers creating virtual buildings with AI include broader users beyond conventional expert designers, encompassing individuals with varying levels of expertise in problem understanding, subject matter, and machine learning~\cite{yang2018grounding}. This innovative designer-AI collaboration, employed by diverse designers, could bring multi-dimensional benefits to the virtual environment.
Firstly, designers could utilize AI to create virtual buildings for efficient and large-scale production, thereby expanding virtual environment scenarios with a variety of content~\cite{jang_automatic_2024}. 
Secondly, designers could prioritize user experiences by considering spatial, social, and physical aspects of human-building interaction (HBI), such as evolving spatial forms and adaptive architectural elements~\cite{10.1145/3309714}.
Thirdly, designers could generate virtual buildings that offer social sustainability and economic benefits by embedding computational frameworks or workflows automatically~\cite{jang_automatic_2024}, aligning with the development goals of the virtual environment. These benefits include time and resource savings, as well as fostering continuous renewal and evolutionary capabilities.
Additionally, deep learning-generated virtual architecture provides convenient access for both experts and inexperienced users in content production through designer-AI collaboration, contributing to the content ecosystem in the virtual environment. One concrete example is that users could create a house or edit land using a user-friendly tool in social VR (Fig.~\ref{fig:VAusecase}n).

Deep learning's powerful capability enables the fusion of virtual and physical realities, facilitating diverse lifestyles and a wide range of user, community, and organizational uses (as shown in Fig.~\ref{fig:VAusecase}).
Virtual buildings could enhance user presence by immersing themselves in vivid, realistic scenes of the virtual environment~\cite{jang_automatic_2024} (Fig. ~\ref{fig:VAusecase}a-c).
Virtual buildings also serve as expansive communication interfaces and containers, fostering social activities and goals in immersive environments~\cite{SCHROEDER2001569}. For instance, users require virtual buildings to hold various activities, events, and communities, such as fan meets (Fig.~\ref{fig:VAusecase}d), club parties (Fig.~\ref{fig:VAusecase}e), and Pride Month parades (Fig.~\ref{fig:VAusecase}f).
Organizations, including companies, cultural institutions, and charity organizations, also require virtual buildings to maintain urban business premises (Fig.~\ref{fig:VAusecase}g-i). Notably, renowned architectural firms have begun designing virtual cities and buildings in virtual worlds (Fig.~\ref{fig:VAusecase}h-k). Furthermore, the interactive nature of virtual architecture facilitates diverse social systems. It delivers sophisticated information regarding social interactive elements in multiple scenarios, as demonstrated by ZAHA's advanced virtual architectural projects incorporating interactive technology and architectural design (Fig.~\ref{fig:VAusecase}h-j).
Lastly, the virtual architecture enables trading, giveaways, and auctions of virtual land as non-fungible tokens (Fig.~\ref{fig:VAusecase}l-m), enriching the metaverse
~\footnote{The metaverse refers to a hypothetical, computer-generated world that integrates with reality through various techniques and concepts~\cite{lee2021all}.}
ecosystem in the future.

The early adoption of emerging technology often encounters accessibility issues, and knowledge from one field may not be easily transferred to another~\cite{DAVIES2018965}. 
In architecture, design research primarily focuses on leveraging neural networks, such as StyleGAN and Pix2Pix, to generate images like floor plans~\cite{chaillou2022advent,chaillou2020archigan}, exterior facades~\cite{mohammad_hybrid_2019, dong_urban_2023}, and stylized images~\cite{del2019church,del2021architecture}. However, there is a lack of effort in collaborating with deep generative models (DGMs) for direct 3D geometry generation. This oversight results in a deficiency of diversified design philosophies, efficient production logic, and precise geometric control in current deep learning-generated architectural designs.
 Despite proposals for computational and algorithmic approaches (e.g.,~\cite{jang_automatic_2024,tencentailab_2023}), design dimensions of virtual architecture have yet to be systematically explored. 
 This gap impedes the advancement of virtual buildings within virtual environments. Our survey aims to bridge this gap by investigating deep-learning-based 3D shape generation methods for virtual architecture.  
We emphasize the importance of considering designer-AI collaboration perspectives, particularly inclusivity, to cater to non-tech-savvy architects and layperson users seeking to create virtual buildings in the virtual environment.

\subsection{Preamble: 3D Virtual Architecture Generated by Deep Learning}
Generated virtual architecture is a vast domain spread over deep learning, computer graphics, and design. 
Therefore, it is necessary to define these terms at the beginning of this survey.

\begin{figure}
    \centering
    \includegraphics[width=\textwidth]{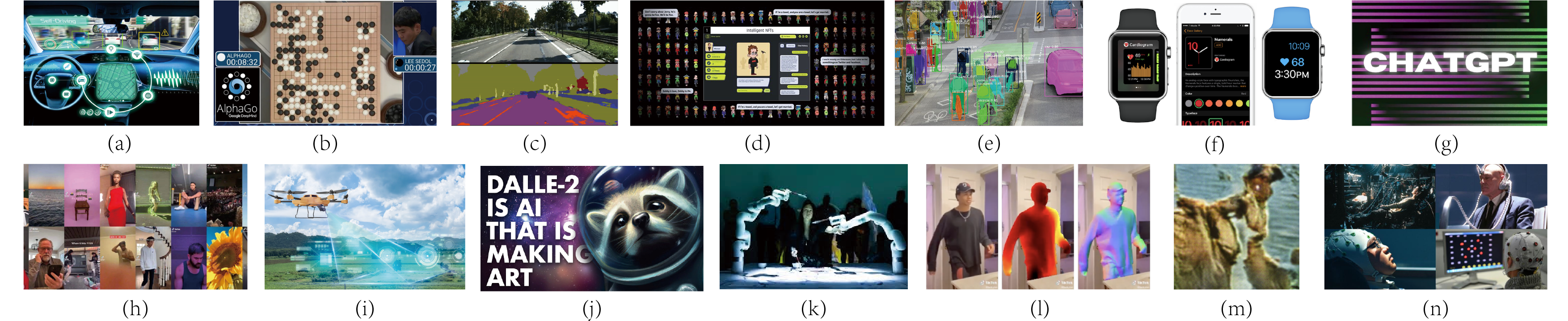}
    \caption
    {
    Applications of deep learning impact our lives in all aspects. (a) Self-drive cars; (b) AlfaGO; (c) Segmentation in the city recognition with computer vision; (d) Mirror World NFT, which shows AI dialogue character with personality and development from learning; (e) OpenCV recognizing the object types in the camera view; (f) Apple watch paired with deep learning detect atrial fibrillation with 97 \% accuracy; (g) ChatGPT developed by OpenAI; (h) Recommendation system in the Tiktok; (i) Smart agriculture implemented by deep learning with drones; (j) DALLE-2, one powerful painting tool empowered by machine learning; (k) D.O.U.G, a collaborative robotic arms interacted with human, learning human behaviors and gestures, performance and created by artist Soug Wen; (l) digital human body generation by 3D reconstruction technique; (m) An AI art movie created by GANs (Casey Reas); (n) BCI (Brain-computer interface).}
    \label{fig:ml-applications}
\end{figure}

\textit{Deep learning.} \textit{Deep learning (DL)}, a subclass of machine learning (ML) and artificial intelligence (AI), has developed rapidly with a boost in data process and computation. 
They rely on paradigms of unsupervised learning. 
As signature deep learning architectures, neural networks such as ANN, CNN, and RNN have played essential roles in manipulating the relationship between the input and output data. The definition of deep learning signifies the system masters the capability of self-learning and experience  enhancement~\cite{surveysarker2021machine}. Deep learning has broad applications across various domains of life. There are plenty of notable examples. Such as the first fully automatic self-driving car, Navlab5 (Fig.~\ref{fig:ml-applications}a); Alpha GO, a computer program that can beat top human Go players (Fig.~\ref{fig:ml-applications}b); Mirror World NFT's intelligent character~\footnote{Mirror Wolrd's official websites: https://link3.to/mirrorworld} that can learn and grow up from human text conversations (Fig.~\ref{fig:ml-applications}d); ChatGPT~\footnote{Introducing ChatGPT, source: \url{https://openai.com/blog/chatgpt}} developed by OpenAI ~\cite{openai_2022}, the ever first intelligent conversational machine model; One of the best recommendation systems in the world (Fig.~\ref{fig:ml-applications}h), which made TikTok stand out from 13.47 million DAUs ~\footnote{A report on TikTok, souce: https://www.statista.com/statistics/1090659/tiktok-dau-worldwide-android/}; DALLE-2, which was commented as the ever-best AI painting tool (Fig.~\ref{fig:ml-applications}j); and the infinite potential BCI (Brain-computer interface) (Fig.~\ref{fig:ml-applications}n); Moreover, the intelligence revolution could not be ongoing without DL, for instance, smart agriculture (Fig.~\ref{fig:ml-applications}i), and smart transportation. Computer Vision relies on deep learning closely, such as notable OpenCV (Fig.~\ref{fig:ml-applications}e), segmentation with vision cognition (Fig.~\ref{fig:ml-applications}c), and 3D scanning techniques (Fig.~\ref{fig:ml-applications}l). Additionally, loads of contemporary digital art were created through deep learning by inputting and processing the data of human gestures and bio-signals (Fig.~\ref{fig:ml-applications}k). Figure 1m represents a cutting-edge example of experimental AI-art films created using GAN. 

\textit{3D Shape Generation Technique.}
It refers to creating digital representations of objects in three spatial dimensions utilizing ~\textit{deep learning models (DGMs)}, which is a foundational task in computer graphics.
Various DGMs like Generative Adversarial Networks (GANs), Variational Autoencoders (VAEs), Flow models, and Diffusion models (DPMs) are employed, each with distinct technical attributes.
For example, 3D-GAN employs probability spaces for generating 3D objects with voxel grid representation~\cite{wu20163DGANlearning}. 
3D shape generation inspires downstream tasks such as object classification, part segmentation, and scene semantic parsing~\cite{harshvardhan2020comprehensive}. 
Moreover, its various innovative approaches, such as descriptive prompts~\cite{guida_multimodal_2023,poole2022dreamfusion,nichol2021glide}, model optimization~\cite{regenwetter2022engineering}, multi-modal interaction~\cite{fu2022shapecrafter,poole2022dreamfusion}, and integrating tools~\cite{poole2022dreamfusion}, collaboratively foster production and applications in design fields representing 3D models as output.

\textit{Deep Learning-Assisted Architectural Form Generation.}
It utilizes neural networks to generate building forms, integrating design features and corresponding purposes~\cite{doi:10.1177/14780771221100102}. 
Its distinctive use of deep generative models (DGMs) sets it apart from other areas in deep learning-assisted architectural design~\cite{doi:10.1177/14780771221100102}, such as environmental performance analysis, and AI-assisted outcome evaluation.
This approach addresses architectural, urban, and environmental challenges by employing 2D images or 3D form generation techniques.
Research in this domain encompasses various aspects, such as generative plans or sections based on spatial features~\cite{ren_spire_2020, asmar2020machinic,yu2020reprogramming, zhang_text--form_2020, zhang20193d}, style transfer for preserving cultural heritage preservation~\cite{ren_spire_2020,liu2021pipes}, and the generation of diverse architectural typologies~\cite{guo2020deep}.

\subsection{Methodology and Related Articles}

\begin{figure}
\centering
\begin{subfigure}{0.4\textwidth}
    \centering
    \includegraphics[height=3.7cm]{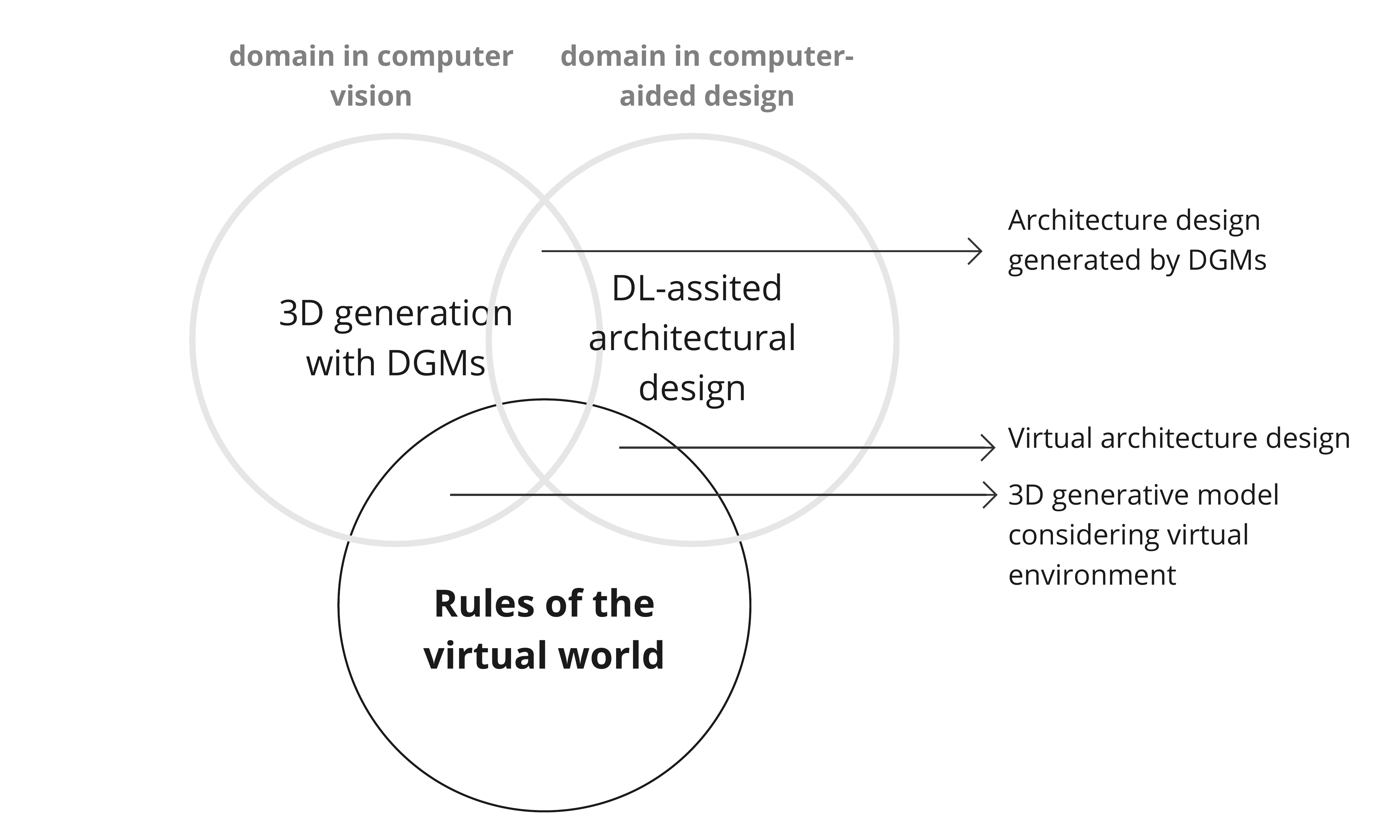}
    \caption{}
    \label{fig:diagram-3circles}
\end{subfigure}
\hfill
\begin{subfigure}{0.54\textwidth}
    \centering
    \includegraphics[height=3.5cm]{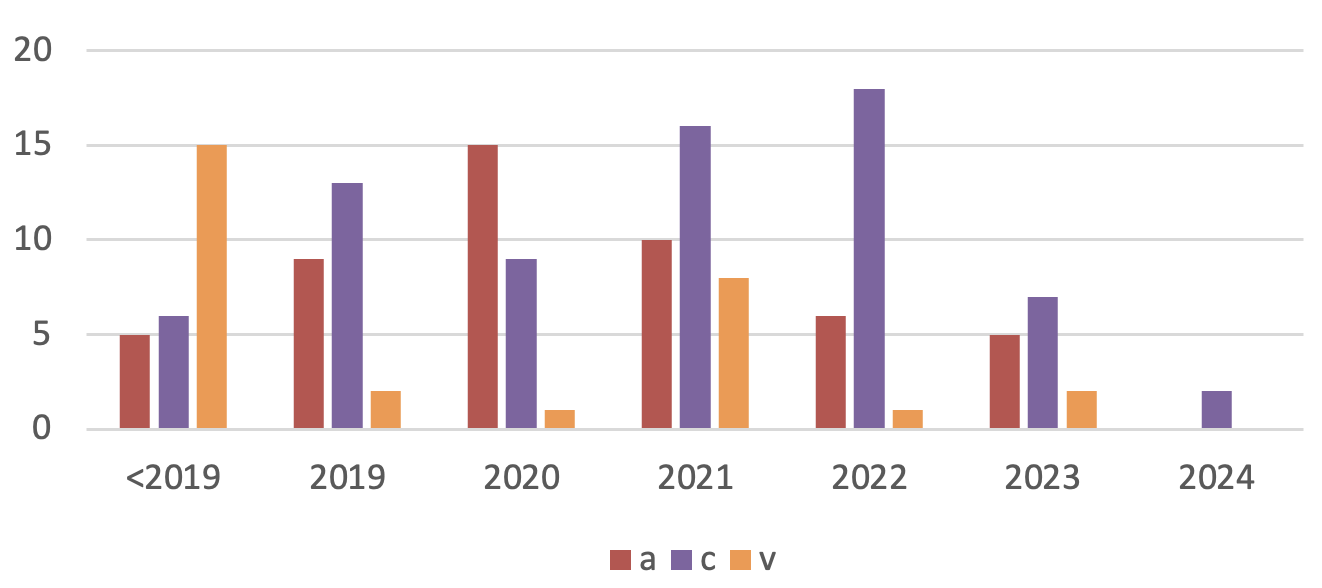}
    \caption{} 
    \label{fig:profile}
\end{subfigure}
\caption{(a) This survey investigates the intersection of various areas. (b) A profile of the number of cited works categorized by different years and topics. The survey's scope and profile of related articles: a -- architectural studies on DGMs; c -- computer vision studies; v -- works on rules in VWs.}
\end{figure}

As for the abovementioned problems, the intersection of 3D generation techniques and virtual architecture is still nearly blank. Therefore, we anchored the three fields to conduct this survey to complement the key insights with each other: 3D shape generation techniques, DL-assisted architectural design, and the design considerations in VWs (see Fig.~\ref{fig:diagram-3circles}).  
This article collects data using the Preferred Reporting Items for Systematic Reviews and Meta-Analyses (PRISMA) framework~\cite{PAGE2021PRISMA}.

\textit{Identification.}
We reviewed a sample of 149 articles and primarily focused on works published between 2019 and 2023 (five years, 81.2\%) as follows: 2024: 2 (1.3\%), 2023: 14 (9.4\%), 2022: 25 (16.8\%), 2021: 34 (22.8\%), 2020: 25 (16.8\%), 2019: 23 (15.4\%), before 2018: 26 (17.5\%) from mentioned three fields (see Fig. ~\ref{fig:profile}). 
We found the articles primarily through publication databases such as ACM Digital Library, IEEE Xplore, ScienceDirect, Springer Link, and CuminCAD. These databases ensure a wide range of computer graphics, virtual environments, and architectural design scope.
We determined our search query, including the primary and second terms as searching keywords, after discussion with inner researchers through multiple sessions of iteration and refinements. These keywords were inspired by previous review papers. The primary keywords in the search query are straightforward, reflecting our survey topic, as follows:

     "artificial intelligence (AI)" OR "machine learning (ML)" OR "deep learning (DL)" OR "Generative Adversarial Network (GAN)" OR "3D-aware synthesis" OR "Variational Autoencoder (VAE)", "Diffusion Model (DPM)"

    AND "architectural design" OR "virtual building" OR "virtual environment" OR "virtual rules" OR "design discipline" OR "building generation" OR "virtual reality (VR)" OR "augmented reality (AR)" OR "metaverse"


We conduct these search queries aligning in publication title, abstract and keyword.
To ensure the subtle and comprehensive searching results in design fields, the second keywords include relating architectural design terms (i.e., AI architecture, design generation, digital architecture, form finding) or technical terms (i.e.,text-to-3D, multimodality, semantic interpretation), as follows:

     "text-to-3D" OR "image-to-3D" OR "image-to-form" OR "language-based" OR "semantics" OR "NLP" OR "multimodal" OR "3D form" OR "form finding" OR "generative model" OR "neural network" OR "Style Transfer"

    AND "computation" OR "human-computer-interaction (HCI)" OR "evaluation metric" OR "perception" OR "bio-signal" OR "emotion" OR "participatory design" OR "aesthetics" OR "real-time interaction"

The final database search was conducted by January 2023.

\textit{Screening, Eligibility, and Inclusion.}
We screened titles and keywords to ensure the inclusion of only full papers and extended abstracts, applying the following exclusion criteria (EC):

\begin{itemize}
    \item [\textbf{EC1}] Excluded articles solely discussing real-world architectural design aspects (e.g., natural environment, BIM, construction, floor plans).
    \item [\textbf{EC2}] Excluded articles not primarily focused on 3D design generation (e.g., analysis of design cognition, limited 2D generative drawing without 3D transition approaches).
    \item [\textbf{EC3}] Excluded articles primarily discussing generative approaches other than deep learning for architecture (e.g., reinforcement learning, genetic algorithms). Only articles discussing deep generative models for building generation were considered.
    \item [\textbf{EC4}] Excluded articles discussing broad architecture design topics (e.g., the architecture of algorithms).
    \item [\textbf{EC5}] Excluded articles related to 3D generation focusing on the classification and retrieval of computer vision approaches.
\end{itemize}

When the keywords and abstracts did not appear as the key information or elements within our investigative scope, we read the entire publication to determine its inclusion. After screening, we selected 111 articles and 19 CV research papers published on arXiv, totaling 130 authoritative articles. 
Additionally, we conducted direct searches through the Google search engine, concluding with 19 articles and 2 relevant architectural projects from the perspectives of virtual worlds, computational architecture, and architectural theory. Ultimately, this survey includes a total of 149 articles and 2 architectural projects.

Various other surveys further locate this scope, as follows: Category 1 (machine learning ~\cite{kreuzberger2023machine, penney2019survey} or deep learning~\cite{hatcher2018survey, kreuzberger2023machine, alom2019state, akinosho2020deep, khan2020survey, alom2019state, sarker2021deep, pouyanfar2018survey, abdar2021review}): 3D shape generation~\cite{surveyshi2022deep, oussidi2018deep, cao2020comprehensive}, scene synthesis ~\cite{xia20223dawaresurvey, zhang2019survey}, applications~\cite{dong2021survey, dargan2020survey}, 3D representation~\cite{guo2020deep}, 3D reconstruction from 2D ~\cite{yuniarti2019review}, and generative models \cite{harshvardhan2020comprehensive, aggarwal2021generative, wang2021generative, jabbar2021survey, pavan2021generative, croitoru2023diffusion, creswell2018generative, harshvardhan2020comprehensive};
Category 2 (artificial intelligence in the architectural design): machine learning-assisted architectural design~\cite{doi:10.1177/14780771221100102,penney2019survey,tamke2018mlarch}, GAN-assisted architectural design~\cite{newton2019generative}, computer-aided architectural design~\cite{stojanovski_rethinking_2022}, historical development~\cite{chaillou2022advent}, infrastructure ~\cite{tamke2018mlarch}, intelligent construction ~\cite{baduge2022artificial, akinosho2020deep,hong2020state}; 
and Category 3 (architecture in a virtual environment or virtual worlds): design disciplinary in virtual reality (theories and applications) ~\cite{bartle2004designingvw}, human perspective in the virtual space ~\cite{lee2021towards,homolja2020impact}, human-building interaction (HBI)~\cite{10.1145/3309714}, and metaverse or virtual worlds ~\cite{bartle2010muds, dionisio20133dmetaverse, lee2021all}.
In contrast, this article reviews the deep learning approaches in the virtual discipline of architecture regarding virtual environment rules, techniques, design principles, and HCI methods. We advocate for an interdisciplinary approach that integrates these research areas. Our survey article uniquely considers the prominent features of the above categories in this topic and further paves the way for AI-architecture integration in virtual environments with the contributions below. 
\begin{enumerate}
    \item 
We provide a comprehensive investigation for the inclusion of DL-assisted architectural generation and deep generative models dedicated to developing a critical lens for computational architecture in virtual environments.
    \item 
We highlight four characteristics of various generation approaches to improving technical understanding and prompting design applications by elaborating on their capabilities and challenges from the whole picture.
    \item 
We propose research agendas and future directions for generated virtual architecture, considering virtual disciplines, deep learning, and architectural design, thus call for interdisciplinary designer-AI collaboration research.
\end{enumerate}

\subsection{Scope and Structure}


While the intersection of deep learning and architecture encompasses various aspects, including design cognition (e.g., feature extraction, image recognition, design evaluation), design generation (e.g., floor plans, rendering, form generation), and assisting tools (e.g., efficiency optimization, strategy prediction), our investigation focuses specifically on articles where deep learning is applied to the generation of 3D virtual architecture, particularly utilizing 3D Deep Generative Models (DGMs).
In this context, two key scope notions warrant emphasis:
Firstly, we underscore 3D DGMs due to their potential to efficiently generate designs aligned with diverse goals, thereby enriching content and affordances in the virtual environment. We also include architectural research that generates 2D images with 3D transitions rather than stopping at 2D-style imaginary drawings.
Secondly, our scope is confined to the generation task, distinguishing it from similar reconstruction and completion tasks, as generation involves actively creating 3D shapes using deep learning. This decision is informed by the prevalent application of generative tasks in architectural design, closely aligning with human cognition.
Furthermore, in line with the virtual context of our inquiry, we prioritize articles where virtual environments can substitute physical ones rather than solely addressing real-world problems like Building Information Modeling (BIM).

The paper begins by reviewing the current problem space in this field, which encompasses the rules of the virtual world, 3D generation techniques, and related architectural design works in Section~\ref{sec:overview}. In the overview of related works, we also emphasize four key focuses derived from the challenges identified in current literature, namely dataset, multimodality, design intuition, and generative framework.
Section~\ref{sec:DGMs} delves into innovative form generation in architecture, exploring generation approaches concerning 3D form transposition and 3D solid form generation. This section discusses four topics: GAN targeting specific training, VAE for specific information extraction, 3D-aware image synthesis, and diffusion model based on conditional text.
Subsequently, Section~\ref{sec:chara} provides a summary of the various characteristics of design generation approaches from different DGMs and design purposes.
Finally, we formulate four grand challenges and three future directions by revisiting the key focuses from the perspective of HCI in Section~\ref{sec:researchagendas}. These challenges encompass agency and communication between design and technique, user consideration, and adaptive tools. We emphasize that innovative methods for automatically generated virtual architecture should prioritize human and social factors.

\section{Overview of Generated Virtual Architecture\label{sec:overview}}
We first define the virtual architecture and its design discipline to clarify our survey and key focus. Then, we briefly introduce deep generative models and 3D representations. Last, we review existing literature on generated virtual architecture.

\subsection{Virtual Architecture and Design Discipline}
Virtual worlds are enduring, computer-generated environments existing online where users interact with others in remote physical locations in real-time, for both work or leisure~\cite{dionisio20133dmetaverse}. 
Research on the virtual world has surged in recent decades, encompassing theory, concept, and technology development as the early stage of the metaverse~\cite{dionisio20133dmetaverse,lee2021all,SCHROEDER2001569,nevelsteen2018virtual}. Two layers form a complete, accessible, and livable metaverse. Firstly, software and hardware architecture technically defines spatial functionality, constraints, and social interaction, which form the politics of virtual worlds~\cite{lessig2009codelaws}. Secondly, immersion, presence, and interactivity as three pillars in virtual worlds further shape user experience~\cite{mutterlein2018three}, constituting the essence of simulating real-life experiences and fostering self-presence, community social activities, and user ownership.

\subsubsection{Virtual Architecture: Mission and Scope}

Virtual architecture is a spatial instance within the virtual world. It is characterized by interactive features related to social attributions and technology frameworks, including realism, ubiquity, interoperability, and scalability~\cite{dionisio20133dmetaverse}. 
The concept of virtual architecture encompasses various buildings and scenes with immersive experiences, such as video games and social VR. 

Our scrutiny reveals an overlapping scope between virtual architecture and human-building interaction (HBI), referring to a consistent goal between the social acceptability of metaverse~\cite{lee2021all} and sustainability of HBI~\cite{10.1145/3309714}. Specifically, HBI focuses on understanding and facilitating human interaction, adaptation, and emotional connection within the built environment~\cite{10.1145/3309714}, whose sustainability highly echos the social acceptability of the metaverse.  Furthermore, researchers have built a close correlation between these two notions relying on user perception in immersive virtual environments aiming at HBI goal~\cite{10.1007/978-3-319-21380-4_32}. 
Additionally, according to conceptualizing HBI by Alavi et al., the built environment comprises social, spatial, and physical aspects~\cite{10.1145/3309714}. 
This work also emphasizes that significant technologies driving HBI include ubiquitous computing, interactive architecture components, and sustainability.
Notably, previous HBI surveys emphasize the inherent attributes of buildings, distinguishing them from traditional HCI~\cite{10.1145/3309714, 10.1007/978-3-319-67684-5_21}.
Therefore, the HBI perspective offers a new approach to understanding the multifunctional nature of virtual buildings~\cite{10.1145/3309714}, regarding encapsulating ubiquitous computing, human-centric interaction, and sustainability goals. We outline these essentials to define the mission and scope of virtual architecture in Table~\ref{tab:disciplines} (Appendix).


\subsubsection{Design Discipline}
Virtual architecture gives rise to new design disciplines distinct from traditional architecture that align with the logic of virtual worlds. The below paragraphs elaborate on the design considerations, including building form, production, and construction methods, as well as the pivotal role of deep learning approaches within these design disciplines.

\begin{figure}

\centering
\begin{subfigure}[t]{0.068\textwidth}
    \includegraphics[height=1.5cm]{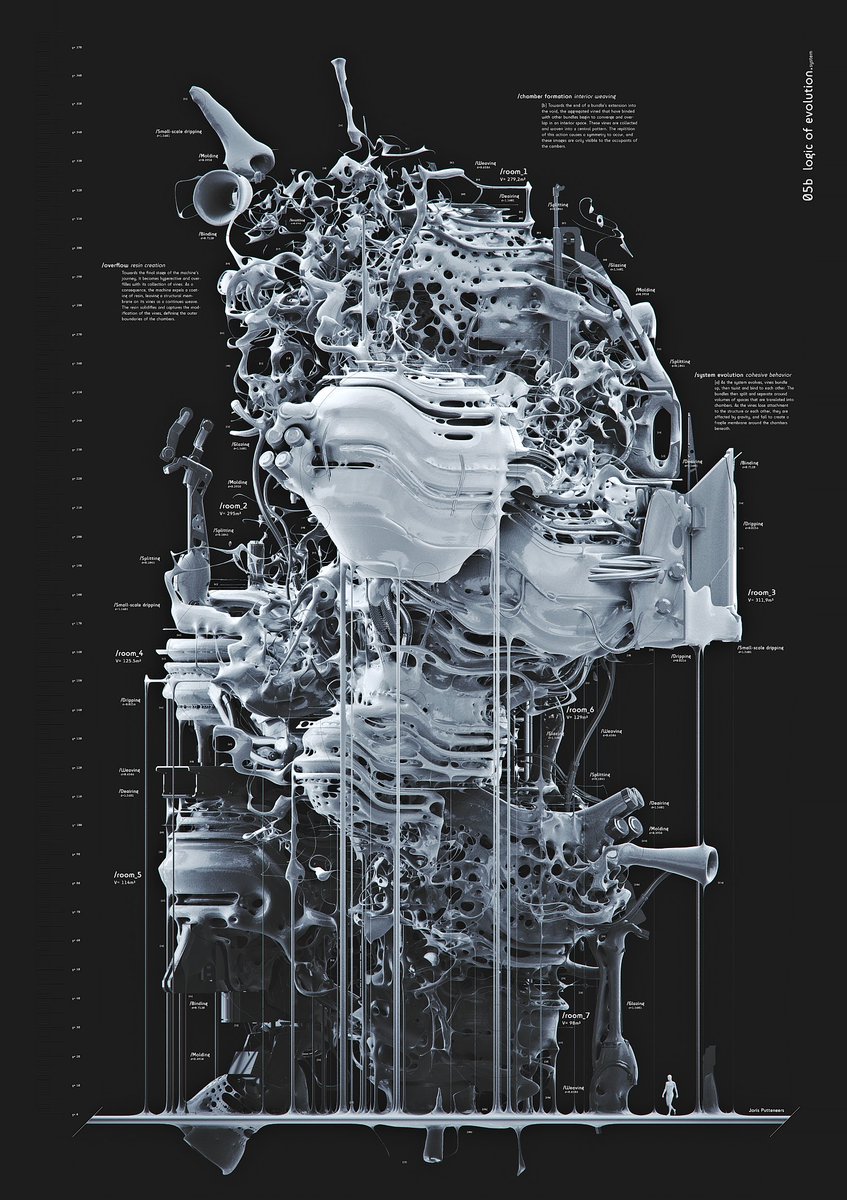}
    \caption{}
    \label{fig:liberty-Synesthesia1}
\end{subfigure}
\begin{subfigure}[t]{0.145\textwidth}
    \includegraphics[height=1.5cm]{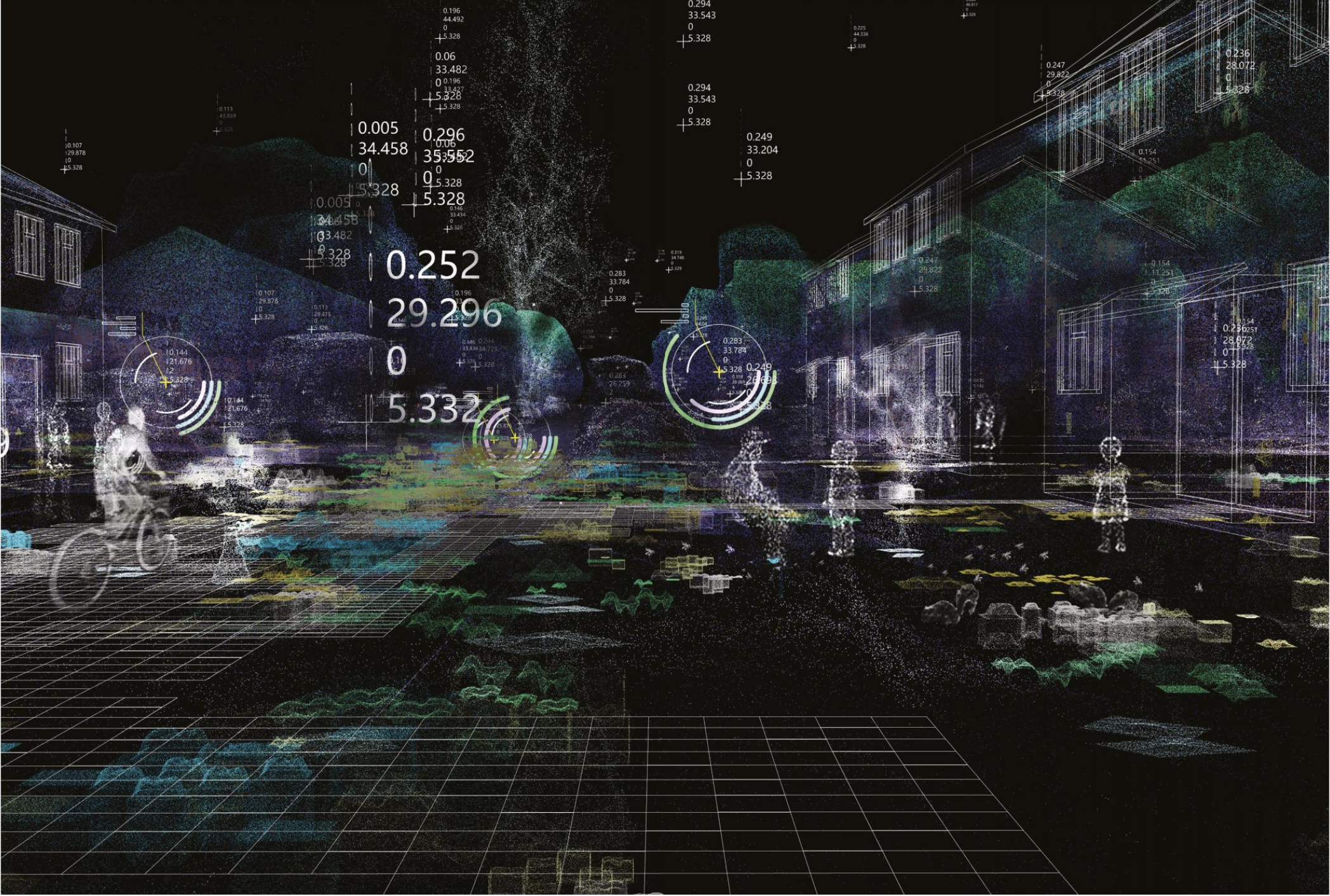}
    \caption{}
    \label{fig:liberty-Emotion}
\end{subfigure}
\begin{subfigure}[t]{0.175\textwidth}
    \includegraphics[height=1.5cm]{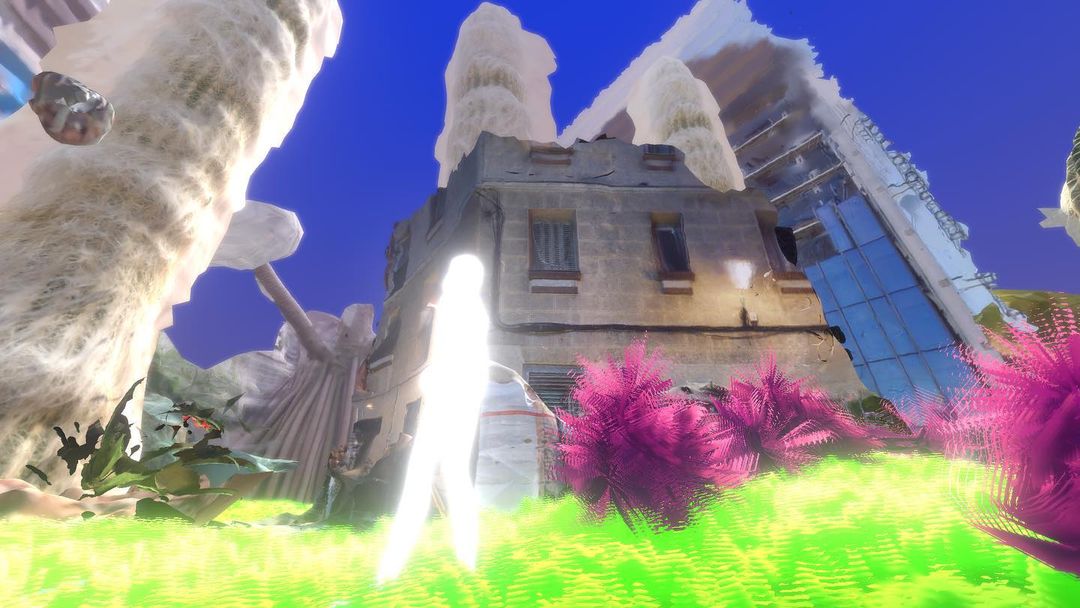}
    \caption{}
    \label{fig:liberty-ISOS}
\end{subfigure}
\begin{subfigure}[t]{0.2\textwidth}
    \centering
    \includegraphics[height=1.5cm]{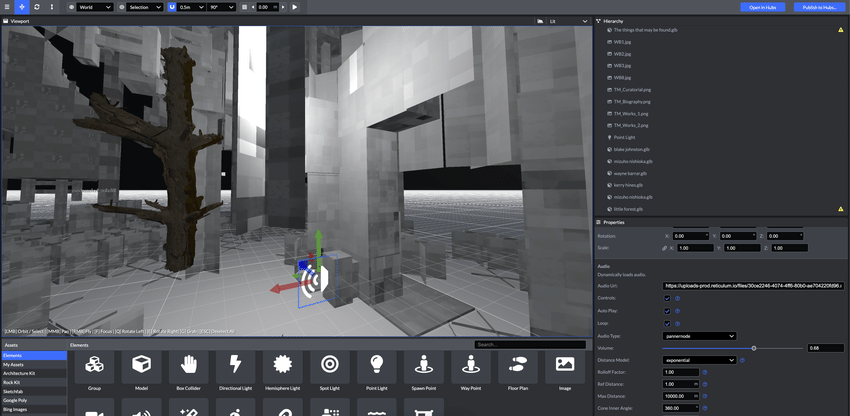}
    \caption{}
    \label{fig:liberty-Forest}
\end{subfigure}
\begin{subfigure}[t]{0.2\textwidth}
    \centering
    \includegraphics[height=1.5cm]{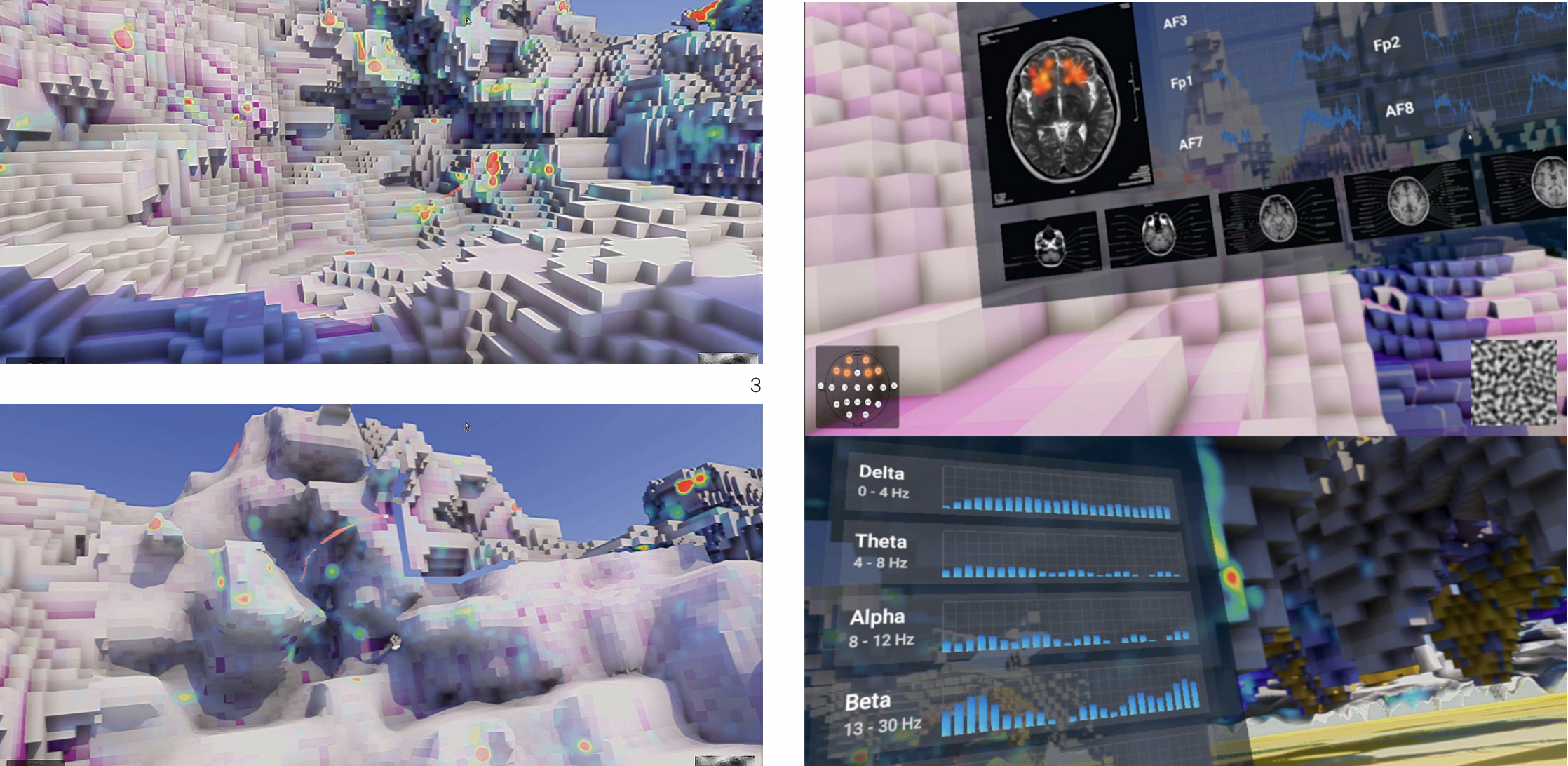}
    \caption{}
    \label{fig:liberty-affective}
\end{subfigure}
\begin{subfigure}[t]{0.12\textwidth}
    \centering
    \includegraphics[height=1.5cm]{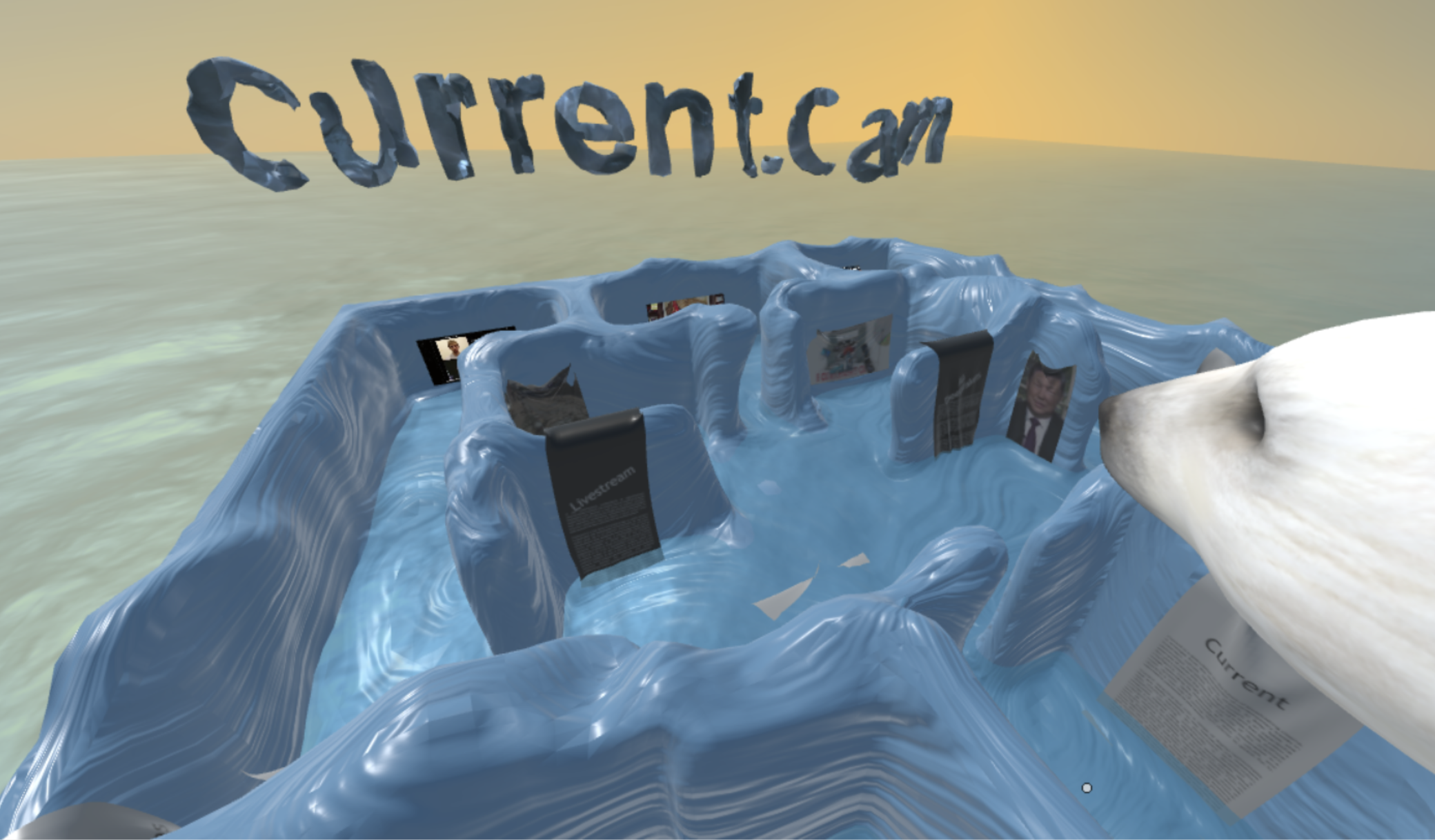}
    \caption{}
    \label{fig:liberty-currentVR}
\end{subfigure}
        
\caption{Virtual architecture projects exhibit various forms:
    (a) Joris Putteneers' "Synesthesia" \protect\footnotemark (2016) creates a surreal and complex architectural construction through algorithmic simulation in Houdini; 
    (b) "E-motion" \protect\footnotemark (2020) by Fei Chen et al. features a digital interface allowing real-time data interaction for rethinking co-living among species; 
    (c) "ISOS" by Viviane Toraci Fiorella, Taza Celilia, and Prandini Alvaro Campo in the "Volumetric Cinema" workshop by Current.CAM (2022) explores the interaction between virtual avatars and dramatic environments; 
    (d) Tane Moleta and Mizuho Nishioka's collaborative project (2021), "Populating Virtual Worlds Together," leads to an autonomous virtual space; 
    (e) Utilizing BCI to capture affective-driven dynamic noise, Barsan et al. (2020) demonstrate 3D volumetric architecture in virtual environments \protect\cite{barsan2020affective}; 
    (f) Current.CAM's VR gallery (2021) features continuous partitioned spaces with fluid blue, reinforcing the shaping of digital interfaces on the human senses.}
\label{fig:liberty}

\end{figure}

\addtocounter{footnote}{-2}
\stepcounter{footnote}\footnotetext{Source:~\protect\url{https://putteneersjoris.xyz/projects/synesthesia/synesthesia.html}}    
\stepcounter{footnote}\footnotetext{Source:~\protect\url{https://bproautumn2020.bartlettarchucl.com/rc18/e-motion}}

 \begin{enumerate}
    \item
    \textbf{Design Consideration.} 
    Deep generative models for architectural generation could easily replace the physical context in current research with a virtual one in the following two aspects.
    First, virtual logic (rendering, computation, graphics, society codes) replaces physical context (i.e., environment, material, economic cost), where users emphasize social gathering activities with unrestricted time and space. 
    Second, virtual logic takes into account more aesthetic, cultural, and human-centered intentions than physical architecture~\cite{barsan2020affective}. 
    Virtual architecture can take this grand responsibility, utilizing deep generative models to incorporate wide definitions and creativity affordances.


    \item
    \textbf{Building Form.} 
    First, deep generative models could form various formats, from solid mesh to the discrete point cloud, to support virtual architecture with flexible and inclusive representations.
    Generation and virtual techniques enable the construction and rendering of cutting-edge buildings in the virtual world by replying to proper datasets, such as Paper Architecture~\footnote{Visionary architecture that could not be built in reality, only as drawings, collages, or models.}, and bionic architecture~\footnote{Bionic architecture is usually computationally adapted to the structure or form of organic matter in nature. Design considerations for biomimetic architecture include living organisms' physiological, behavioral, and structural adaptations.}, which previously not be implemented in reality.
    Second, leveraging deep generative models could form a diversified and rich shared online space since this technique not only incorporates the vast and various datasets contributed by users but also flexibly adjusts its framework or workflow. 
    Space geometry has been expanded to space intelligence, representing the generative logic of forms transformatively. For instance, space form could be embedded in technical capabilities as assistance, such as algorithmic fluid form (Fig.~\ref{fig:liberty-Synesthesia1},~\ref{fig:liberty-currentVR}), spatial data with point cloud (Fig.~\ref{fig:liberty-Emotion}), interactive narrative (Fig.~\ref{fig:liberty-ISOS}), collaborative co-constructing (Fig.~\ref{fig:liberty-Forest}), and creative voxel girds output (Fig.~\ref{fig:liberty-affective}).
        

    \item 
    \textbf{Production Mode.} Undoubtedly, deep learning with 3D generation capability facilitates novel production modes. First, the construction mode has changed into modeling form from 3D space with modeling software assisted instead of the traditional floor plan of function layout in architectural design.
    Second, the process of designing and building virtual architecture has changed. ZHVR group illustrates this new construction process as five phases: concept design, developed design, technical design, construction, and post-construction~\footnote{Reference the diagram named Redefinition of AEC project ecosystem by ZHVR}. Meanwhile, the stakeholders have changed in this process; architects and programmers have become architects and engineers.
 \end{enumerate}

By highlighting these key points, we craft a virtual architecture guideline urging a broad understanding grounded in knowledge and expertise for virtual buildings.




\subsection{Deep Generative Models and 3D Representations}
We briefly overview the progression of deep generative models for 3D representation, including 3D shape generation and 3D aware image generation. 

\begin{figure}
    \centering
    \includegraphics[width=1\textwidth]{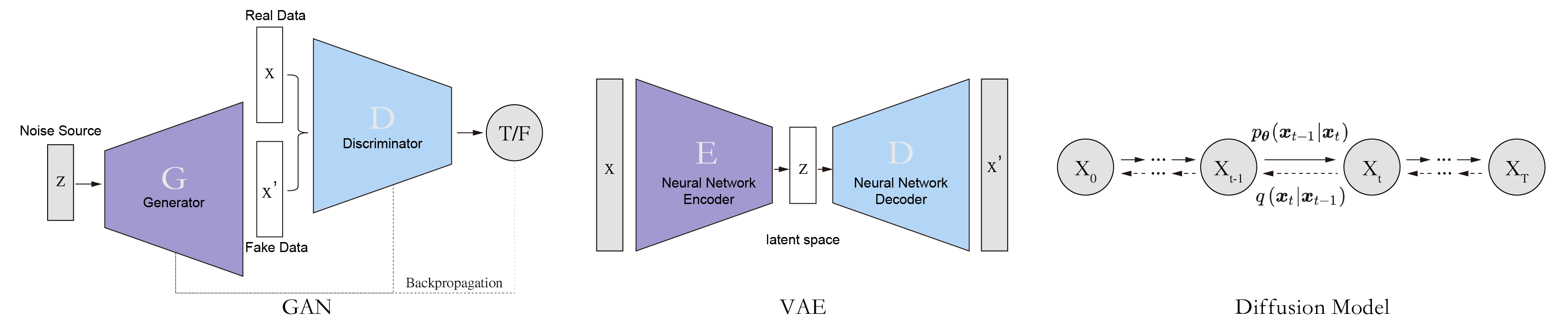}
    \caption{The frameworks of GANs, VAEs, and Diffusion Models.}
    \label{generativemodels}
\end{figure}

\subsubsection{3D Shape Generation} Traditional deep generative models contribute to 3D shape generation alongside well-known models such as Generative Adversarial Networks (GANs) and variational autoencoders (VAEs). Other models like normalizing flows (N-Flows), recent diffusion probabilistic models (DDPMs), and energy-based models (EBMs) have also made significant contributions by learning from the similarity of given data. These deep generative models generate a tangible 3D object ready for rendering. It conveys a latent variable to a high-quality image. 
Although each model has benefits and has shown remarkable progress in recent years, the architecture domain primarily relies on GANs.
In contrast, VAEs and the latest diffusion models are designed for a few research. Considering the relevance, we thus introduce the GANs, VAEs, and diffusion models in detail rather than an exhaustive list of models included in other CV survey articles (Fig. ~\ref{generativemodels}).

\textit{GANs.}
GANs are a type of semi-supervised learning relying on the noise value. The GANs rely on machine learning algorithms to construct two neural networks: 
one is a generator, and another is a discriminator. 
It trains a large database employing a zero-sum game between two neural networks to generate agnostic creative results.


\textit{Variational Autoencoders.} Variational autoencoders are probabilistic generative models that employ neural networks, utilizing both encoders and decoders alongside inputs and outputs.
The latent space of a VAE learns data features and simplifies data representations to aid model training for a specific purpose. 
During training, regularization ensures that the latent space of a VAE possesses acceptable qualities and can generate new data~\cite{Kingma2019AnIT}. 
Furthermore, the term "variational" originates from the tight connection between regularization and the variational inference technique used in statistical analysis. 

\textit{Diffusion Models.} By modeling the dispersion of data points in latent space, we can uncover the underlying structure of image or volumetric datasets, exemplified by Denoising Diffusion Probabilistic Models~\cite{ho2020denoising}.
This entails training a neural network to eliminate the blurring effect of Gaussian noise on an image, resulting in the significant advantage of producing sharp and detailed features. 

\subsubsection{3D-Aware Image Synthesis} This approach extracts latent vectors from the latent space and decodes them into a target representation by using GAN.
Generally, the generation pipeline is for an image with 3D awareness as a result, and it also starts with an image as a generative source. 

\subsubsection{3D Representations}

\begin{figure}
    \centering
    \includegraphics[width=.8\textwidth]{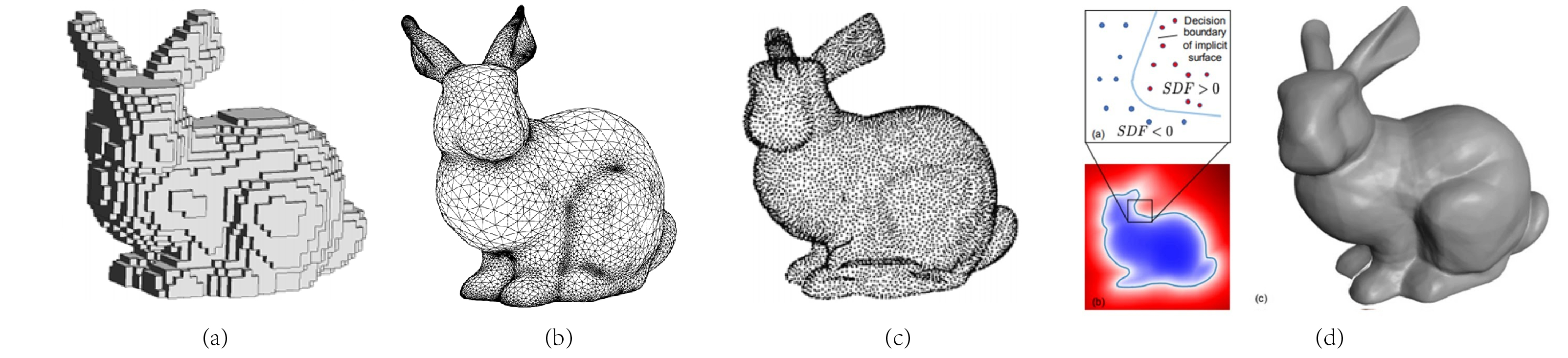}
    \caption{The figure illustrates various 3D representations commonly used in generated architecture: (a) Voxel grids, (b) Meshes, (c) Point clouds, and (d) Neural fields.}
    \label{format}
\end{figure}

These two types of 3D generation yield diverse representations of 3D scenes in computer vision and computer graphics.
The 3D representation in 3D shape generally includes explicit representations, such as voxel grids, point clouds, meshes, and implicit neural fields. A 3D-aware image includes depth or normal maps, voxel grids, neural fields, and hybrid representations. The integration between them and the architecture generation also varies.
For example, a point cloud is often considered an input source for training generative models.
The 3D representation is articulated in existing survey research as a classification (Fig. ~\ref{format}). Below are brief descriptions.

 Architectural design prefers explicit representations due to the controllability, familiarity, visualization, and availability regarding modifying in 3D modeling software. 
Explicit geometric representations are more accessible to visualize and interpret as they directly represent 3D space. The designers can precisely position and adjust each point or voxel, allowing for more accurate control over the shape and form of the generated geometry.
Nevertheless, implicit representations (neural fields) have considerable possibilities in architectural research regarding their benefits to offer more flexible, continuous, and efficient representations of geometry.

\textit{Voxel grids.} It refers to a three-dimensional grid of values organized into rows and columns. The grid contains rows, columns, and layer intersections, referred to as a voxel, i.e., a miniature 3D cube \cite{Deng2020VoxelRT}.

\textit{Point clouds.} A point cloud is a distinct collection of data points in space~\cite{Rusu20113DIH}, which might indicate a three-dimensional form or item through Cartesian coordinates (X, Y, Z) assigned to each point location.

\textit{Meshes. } A 3D mesh is the polygonal framework upon which a 3D object is built \cite{nichol2021glide}. Reference points along the X, Y, and Z axes describe the height, breadth, and depth of a 3D mesh's constituent forms. It is important to note that creating a photorealistic 3D model sometimes requires many polygons.

\textit{Neural fields. } It creates images by using traditional volume rendering methods to query 5D coordinates along camera rays and projects the resulting colors and densities onto a 2D plane. Despite its use of depth data, the scene geometry is rendered in exquisite detail, complete with intricate occlusions \cite{Mildenhall2020NeRFRS}.

\textit{Hybrid representation.} This refers to a hybrid pipeline of 3D representation for the pre-training in a 3D feature space embedded in both the virtual and actual worlds. The hybrid pipeline can include multitudinous data sources and image frame features \cite{shen2021dmtet}, depending on the generation purposes of the 3D volumetry.

\subsection{Generated Virtual Architecture with Deep Learning\label{ad-ml}}
Generating virtual buildings is an urgent and timely topic for virtual architecture since the traditional design process has shifted due to AI intervention.
As our scope mentioned, we only investigate design generation through deep learning in the following sections. We summarize the current six approaches of 3D generation according to design process modality from both 3D transposition and 3D solid form generation (see Fig.~\ref{fig:ML_AD-methods}).

\subsubsection{Related Works} 
Integrating deep learning into architectural design has significantly increased in recent years. Initially, neural networks in deep learning were widely utilized for style transfer in 2D drawings of architectural design (depicted as \protect\textit{3D transposition} in Fig.~\protect\ref{fig:ML_AD-methods}), such as floor plans~\cite{chaillou2022advent,chaillou2020archigan}, sections~\cite{yu2020reprogramming}, and facades~\cite{mohammad_hybrid_2019, dong_urban_2023}.
Most approaches employ a zero-sum game of these neural networks to blend architectural styles with design intentions and then manually assemble these images into 3D buildings as a post-processing step~\cite{yuan_serlio_2022}. However, this indirect method poses significant challenges for designers due to the absence of interpretative embeddings in each image, such as materiality, style, proportions, or program representation~\cite{guida_multimodal_2023}.
Additionally, a general lack of knowledge and expertise in generation techniques regarding frameworks and algorithms in the architectural field leads to underexploration in the design.


Recent works have shifted focus from 2D images to 3D shapes, utilizing techniques such as 3D GANs, VAEs, and 3D diffusion models (depicted as \textit{3D solid form generation} in Fig.~\ref{fig:ML_AD-methods}). 
Current approaches support architectural generation from different modality inputs, such as text, image, and 3D, leveraging different model features. Generally, GAN utilize labeled 3D datasets according to design meaning, such as architectural elements. VAEs generate 3D models utilizing reduced 3D features recorded in latent space with an encoder-decoder. Although the application of 3D diffusion models in architectural design is limited, their multi-modal integration capability and diverse methods suggest significant potential for future use.
Additionally, delving into 3D generation requires a higher technical threshold than simply applying generative models. For instance, in the generation process, GANs require careful adjustment of hyperparameters for stability, network depth, width, and kernel size to ensure performance and output quality. Additionally, considerations of computational resources, dataset characteristics, and random noise input to the generator are essential for effective training~\cite{mueller_3d_2023}.

Both 2D and 3D methodologies shed light on the evolving trend towards designer-assisted AI workflows rather than mere AI-assisted design. Shi et al. underscored the importance of regulation and training by designers in such designer-AI collaboration, alongside ideation, creation, visualization, and testing~\cite{shi2023understanding}. They outline four possibilities regarding regulation and training for better aligning with design goals: curating training datasets, providing semantic information, regulating AI's roles and behaviors, and assisting AI in improving performance~\cite{shi2023understanding}. 
Architectural design experiments have also propelled such pioneering endeavors. For example, Huang proposed an experimental approach in semantic analytics through human interpretation~\cite{jeffrey_huang_gans_2021}, facilitating mutual learning between humans and AI to generate cultural and architectural meaning. Overall, this shift enables designers to contribute to future architecture in virtual and physical realms by transitioning from traditional design roles to "technical design."

We identified four key focuses by thoroughly examining the literature on architectural design employing deep learning: dataset, multimodality, design intuition, and generative framework (see Table~\ref{tab:keyfocus}, Appendix). Notably, our focus is specifically on architectural design and generation considering 3D form, aligning with the scope of our survey.

    \subsubsection{Dataset} Dataset preparation comprises data collection, data cleaning, and filtering, which is the first and foremost step in deep learning-assisted architectural design. However, 3D building datasets regarding specialized design and general applicability are currently scarce and underexplored. Despite being scrutinized, we found only three building datasets that can aid in 3D form generation~\cite{ondrej_vesely_building_2022, mueller_3d_2023, peters2022}. Selvaraju et al. first introduced BuildingNet, a large-scale collection of 3D building models with labeled exteriors~\cite{selvaraju2021buildingnet}. Building upon this work, Mueller developed the HouseNet dataset, a selective collection of preprocessed models, which was subsequently utilized in her design work~\cite{mueller_3d_2023}. Besides, Peters et al. introduced the 3DBAG dataset, containing 3D building models of the Netherlands~\cite{peters2022}, featuring multiple levels of detail. 
    The 3DBAG dataset was further leveraged by Vesely in deep learning-assisted architectural design utilizing heatmap data for generating 3D geometry building~\cite{ondrej_vesely_building_2022}, showcasing its potential in design.

    Therefore, architectural design research has to prioritize building 3D datasets through manual pre-processing before the generation process. This includes generating models using various tools, such as generating massive models approach~\cite{sebestyen_towards_2021}, scanning with specialized scanners~\cite{ccakmak2022extending}, segmenting architectural elements~\cite{ccakmak2022extending}, and formatting and transforming data~\cite{mueller_3d_2023}. The lack of adequate and unified datasets hampers further exploration of broad user affordances and democratized tool development.

    \subsubsection{Multimodality}
\begin{figure}[t]
\centering
    \includegraphics[height=6.4cm]{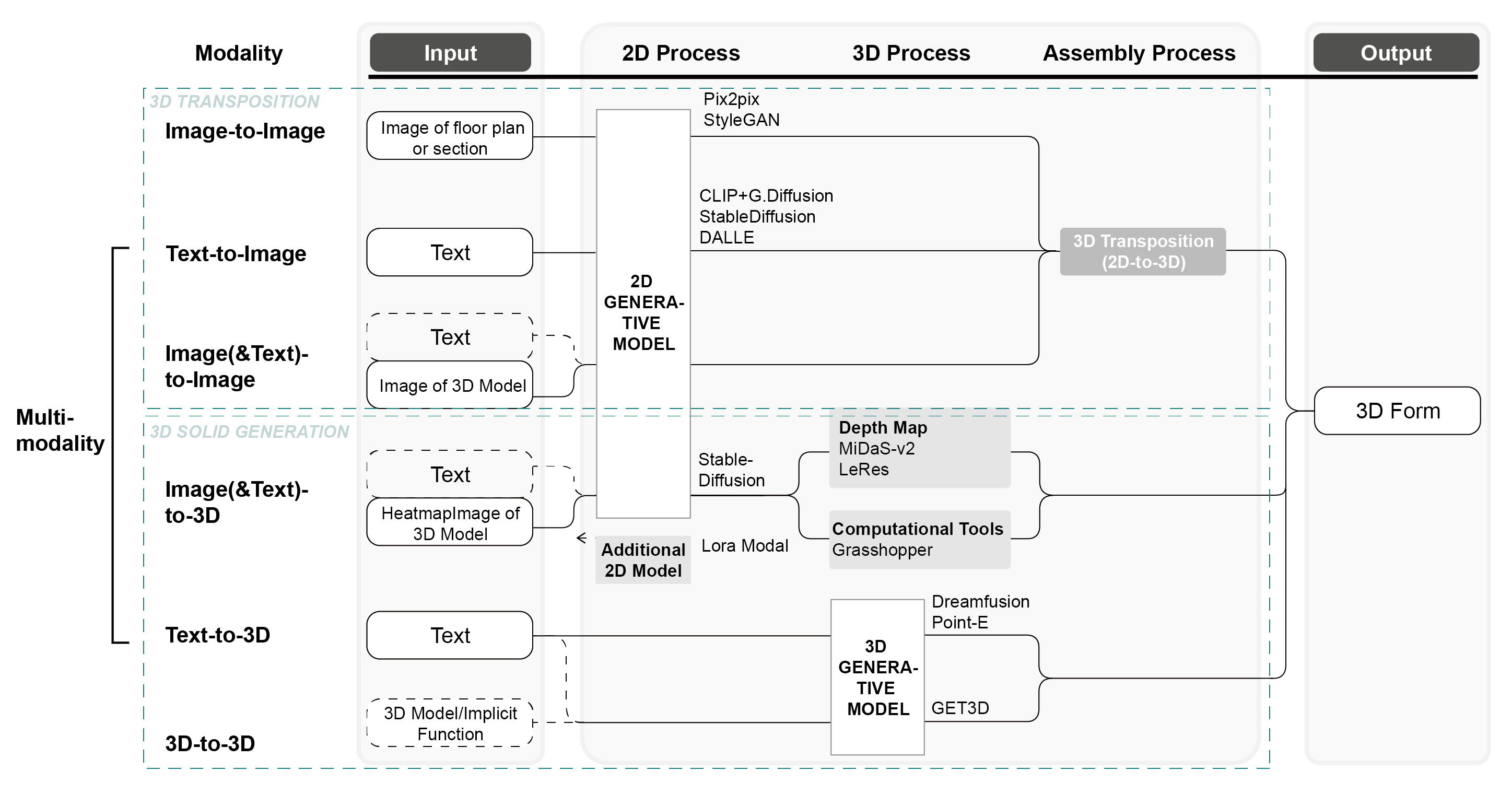}
    \caption[Various architecture design methods by deep learning towards multimodality, then divided into two groups, namely 3D transposition and 3D solid form.]
    {3D transposition leverages 2D generative images to transpose into 3D form in an additional manual or computational way.
    In contrast, 3D solid form generation leverages 3D deep generative models, including 3D GANs, VAEs, and 3D diffusions, to generate 3D form directly.
}
    \label{fig:ML_AD-methods}
\end{figure}

    Multimodality leverages natural human communication capabilities by integrating multiple modalities, such as speech, gestures, and visuals, within a single channel.
    The current literature underscores the significance of multimodality in deep learning methods, particularly in utilizing human understanding as input, such as descriptive prompts and stylized images. 
    Generally, these diverse design methods can be classified as follows, which can also be divided into two groups: 2D and 3D generative approaches with deep learning (\textit{3D transposition} and \textit{3D solid generation} as shown in Fig.~\ref{fig:ML_AD-methods}):

        \textbf{Text-to-Image with Understanding and Interpretation.} This method utilizes generative images from text inputs, such as keywords or descriptive languages, then manually translates images to 3D form by designers with design understanding and interpretation. This process could generate more accurate and meaningful architecture synthesis from images. It could also expand existing design cognition by style transfer,  such as imaginary visual representation.
In addition, generative images in architectural design usually inspire to aid ideation, form finding, and result rendering. 
However, advanced integrated models only rely on non-architectural datasets, such as DALLE2, Stable Diffusion, and CLIP integrated with G. Diffusion, which leads to stiff outcomes lacking design intention. 
For instance, the generative images fail to interpret further architectural meanings, such as "ecological architecture" or "parametricism"~\cite{guida_multimodal_2023}.

        \textbf{Text/Image-to-Image with Vision and Sights.} This method combines visual assembly of inputting 3D models as an image instance with the \textit{text-to-image} method to render buildings. It only renders images as reference outcomes, not implementing 3D models directly. 
        This method enables a quantitative understanding of architecture and grants designers greater agency~\cite{guida_multimodal_2023}. Designers manually convert architectural context (e.g., massing areas, building code, site constraints, and program distribution) into 3D form images as input. Although currently employed in real contexts, its applicability in virtual architecture as 3D form input is acknowledged regarding context and norms. In this method, different models offer designers varying flexibility in adjustments. For instance, Stable Diffusion permits adjusting model parameters and image settings (e.g., image size, clip guidance scales, seed), while DALLE2 does not.

        \textbf{Text-to-Image-to-3D.} 
        This method conducts a pipeline combining \textit{text-to-image} with high-automatic 3D computation by extracting 2D depth information from generative images. Such 3D computation approaches encompass various approaches, such as NeRF, LeRes, and Grasshopper. Current works display the generation of building interiors and exteriors considering various conditions. The latest research developed an integrating tool hosted in Grasshopper to empower architectural design to integrate this mixed approach~\cite{guida_multimodal_2023}.

        \textbf{Text-to-3D with Language.} Advanced deep learning models enable 3D shape generation through descriptive language input, such as 3D Diffusion and Point-E. However, this method has not been widely utilized in architectural design due to a lack of architectural element detail, as our investigation revealed. We delve deeper into this topic in Section~\ref{sec:DGMs}.

    \subsubsection{Design Intuition} 
Design intuition refers to processing generative techniques to cooperate with intended design goals \cite{koh20223dganhousing, ccakmak2022extending}, determining the architectural meaning of design outputs through extended interpretability.
Although many factors can affect the design results aiming to enhance design intuition, we list the two most mainstream ones:
Generative images could be tailored to the design purpose in \textit{3D transposition} approaches by incorporating the desired style. For instance, Ren et al. utilized a StyleGAN to blend the Gothic architectural style with an original plan by inputting the real essence of a Gothic-style floor plan \cite{ren_spire_2020}. 

Preprocessing datasets could also facilitate the design intuition in both\textit{ 3D transposition} and \textit{3D solid form approaches} \cite{koh20223dganhousing, ccakmak2022extending}, which ensures the dataset contains the necessary and accurate architectural elements imbued with "design meaning." 
However, the current approach often involves preprocessing data by designers according to specific design requirements, demanding significant human and computational effort to drive interpretability toward design objectives. 
For example, Çakmak conducted architectural element segmentation on scanned point clouds to generate meaningful components with a GAN and an encoder-decoder architecture \cite{ccakmak2022extending}.

    \subsubsection{Generative Framework}
     This refers to adapting generative frameworks regarding workflows and generative models to correspond to specific design objectives. 
    Rather than simply applying existing models, some recent research endeavors are integrating, adjusting, and developing frameworks. For example, Kahraman et al. investigated the post-integration workflow of a flexible, customized auxiliary discriminator network to enable multi-objective control of generation, catering to the combination of multiple abstract criteria for design purposes \cite{kahraman_augmenting_2021}. 
    
    Furthermore, some studies have shown promise within generative frameworks in connecting 3D geometries and algorithmic manipulation, as well as merging geometry utilizing latent space~\cite{miguel_deep_2019, sebestyen_towards_2021}. This exploration involves investigating hyperparameters and loss functions in GANs while variable manipulation in the latent space of Variational Autoencoders (VAEs). For instance, Mueller examined various hyperparameter options and combinations in GANs to understand their impact on generating meaningful single-family home buildings \cite{mueller_3d_2023}. In contrast, Sebestyen et al. delved a VAE deeper into manipulating geometry by adjusting variables in latent space and incorporating semantic information to achieve intentional output results~\cite{sebestyen_towards_2021}.

\section{Generated 3D Architecture: A Paradigm Shift\label{sec:DGMs}}
\begin{figure}
    \centering
    \includegraphics[width=\textwidth]{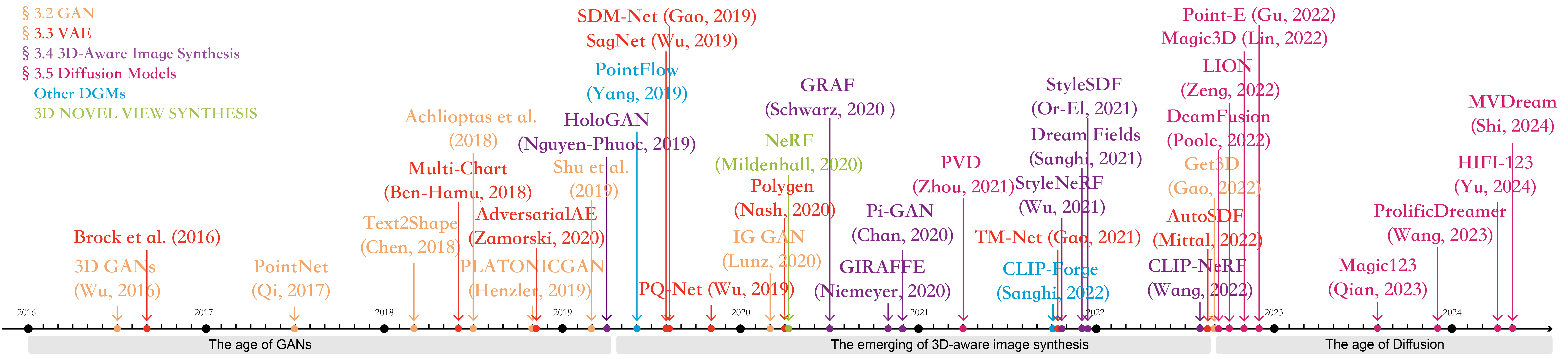}
    \caption{A systematic taxonomy for reviewing generation approaches on virtual architecture design with DGMs.}
    \label{fig:taxonomy}
\end{figure}

To examine the gap between the two fields and the potential advantages of each approach for virtual architecture, we investigate various generative approaches using different DGMs across the fields of CV (Fig. ~\protect\ref{fig:taxonomy}) (Table~\protect\ref{tab:CV3dshape} and Table~\protect\ref{tab:3Dawareimage}, Appendix) and architecture (Table ~\protect\ref{tab:arch}, Appendix). 
Since the existing related architectural works do not systematically form solid design philosophies, we elaborate on these works following their emphasized generative approaches, including 3D transposition (Section~\ref{sec:2D-3DGAN}), generation by GANs (Section~\ref{3DGAN}), VAEs (Section~\ref{3DVAE}), 3D-aware image synthesis (Section~\ref{3DAware}), and diffusion models (Section~\ref{3DDPMs}). 
The first approach relies on additional transposition from generative images, while the others generate geometry directly.



\subsection{Constrained Approaches Using 3D Form Transposition \label{sec:2D-3DGAN}}

In the past, DGMs have primarily focused on 2D image processing tasks, such as texture synthesis~\cite{gao2021tm}, style transfer~\cite{karras2019style}, photorealistic image generation~\cite{nichol2021glide}, and text-to-image synthesis~\cite{nichol2021glide}. Architectural research has similarly centered around 2D to 3D transposition using image-to-image translation networks like Style Transfer \cite{ozel2020interdisciplinary, ren_spire_2020, liu2021pipes, zhang20193d}, StyleGAN \cite{del2019church, zhang20193d, zhang2021machine,karras2019style}, Pix2Pix GAN \cite{yu2020reprogramming, di_carlo_generating_2022}, and CycleGAN~\cite{dong_urban_2023}.

\subsubsection{3D Form Transposition}




This approach converts generative images into 3D representations through computational calculations or manual adjustments. The process starts by segmenting a 3D model into discrete images, such as sections, plans, and projections from various viewpoints. 
For example, Zhang and Blasetti combined two architectural forms using generated 2D-pixel images through discrete section slicing and multiple views \cite{zhang20193d}. Compared to direct 3D approaches, this method offers easier usability and higher computational efficiency as it is based on generative images.

Research in this approach utilize traditional architectural design principles to manipulate spatial features in 2D images on the plan~\cite{ondrej_vesely_building_2022, asmar2020machinic}, section~\cite{bank2022learning}, facade~\cite{del_campo_deep_2022,campo2022generali}, and perspective images~\cite{jeffrey_huang_gans_2021, kim2022perspectival}. For instance, Asmar and Sareen employed a StyleGAN to leverage latent space for image generation, followed by 3D voxelization using vector arithmetic, with a specific focus on generating a single building~\cite{asmar2020machinic}. 
Vesely also utilized Pix2Pix GANs while generating building massings within urban environments by training on high-fidelity 3D building models~\footnote{The dataset is 3D BAD. An overview of 3D BAG, source: ~\url{https://docs.3dbag.nl/en/}},
and incorporating multi-layered site context information~\cite{ondrej_vesely_building_2022}.
Whereas Bank et al. experimentally developed building forms with 3D point clouds from sampling generative images aiming for teaching methods~\cite{bank2022learning}. 
In contrast, some research utilize pixel projection to convert 2D-pixel values, generated by latent walks, into a corresponding 3D model.~\cite{campo2022generali,del_campo_deep_2022}. In these works, Del Campo et al. innovatively generated a house emancipating from a canonical AI-generative approach by intentionally challenging curve-fitting techniques and embracing data scarcity or overfitting~\cite{del_campo_deep_2022}. 
In contrast to prior approaches, some researchers utilize latent space for downscaled perspective information storage, condensing 3D data into 2D images for manipulation. 
For example, Huang et al. generated rendering results with various perspectives by interpolating images generated from the latent space of GAN~\cite{jeffrey_huang_gans_2021}. 
This approach extends prior research by incorporating a rotation of latent space for 3D perspective drawing~\cite{kim2022perspectival}.
This reflects a critical focus on spatial perception in DL-assisted generation, following architectural principles, and interpreting scale from various perspectives.

The 3D transposition approach exemplifies how freedom of computation and extensive manipulation shape the future of virtual architecture with its focus on intricate variations. Given that generative images are lightweight and easy to process, this method ensures high-resolution retention in both input and output. 
However, this method sacrifices geometry controllability and production efficiency for a generation. Additionally, significant human labor is required to establish an automatic workflow. These shortcomings have urged a paradigm shift in generative approaches towards direct geometry production.


\subsection{Approaches Using GANs\label{3DGAN}}
GANs generate 3D form through training distributions by playing zero-sum games with uncertain but high-quality output. Thus, they lead to typical applications in testing the agencies of AI,  opting for conceptual design, and democratizing design.

\subsubsection{3D Shape Generation with GANs}
GANs have emerged as a 3D generation method from a probabilistic space in convolution networks, 
capable of producing various explicit representations, including point clouds \cite{cai2020learning} or voxel grids \cite{michalkiewicz2019deep, mescheder2019occupancy, dinh2014nice, gulrajani2017improved, litany2018deformable, lorensen1987marching, nijkamp2020anatomy, genova2019learning, lunz2020inverse}, as well as implicit neural functions like occupancy fields and signed distance functions (SDF). Wu et al. employed a generative adversarial network architecture to generate 3D voxel grids by capturing the probability distribution of 3D shapes \cite{wu20163DGANlearning}. Although approaches like PLATONICGAN and IG GAN also produce 3D voxel grid models from unstructured 2D image data using GAN~\cite{henzler2019escaping, lunz2020inverse}, the drawback of this method for voxel grids is the inability to achieve fine-grained voxels due to the cubic increase in computational cost. Additionally, point cloud representation, obtained as raw data through depth scanning, presents various challenges in GAN-based generation, including convergence issues, utilization of local contexts, and high memory consumption \cite{achlioptas2018learning, shu20193d, hui2020progressive, arshad2020progressive, ramasinghe2020spectral}. Mesh representation, commonly used in 3D modeling software, faces difficulties in applying deep generation models due to non-Euclidean data and challenges in connecting mesh vertices to compose shapes \cite{surveyshi2022deep}. While methods like the multi-chart approach address irregular mesh structures~\cite{ben2018multi}, approaches like Get3D enable high-quality geometry and texture from 2D image collections by incorporating differentiable surface modeling and rendering into GANs \cite{gao2022get3d}.

\subsubsection{3D GAN Utilized in Architecture}


This refers to generating 3D solid form through a direct 3D data acquisition, evaluation, transformation, and rearrangement using deep generative models ~\cite{steinfeld2019fresh}. 
This approach, aligning training datasets with spatial design parameters, usually achieves specific design objectives.
Specifically, designers utilize datasets with design meaning by labels or semantic interpretation. 
For instance, Immanuel Koh employed 3D GAN to generate massive Singapore high-rise buildings by labeling building exteriors and interiors of training models~\protect\cite{koh20223dganhousing}.
Additionally, researchers and designers have fully probed dataset forms related to architectural structures to address the deficiency in generating precise geometries. 
Zheng and Yuan innovated training coordinate vectors of building surfaces in neural networks to find forms based on NURBS~\footnote{Non-uniform rational basis spline, a mathematical model using basis splines to represent complex curves, surfaces, and solid form.} with different design features~\cite{zheng2021generative}.

Innovating generative methods using 3D GANs for fidelity and accuracy remains a novel and timely topic in AI-assisted architectural design. Some researchers integrate additional computational methods or generative models with 3D GANs to expand the range of design cases for 3D form generation. For instance, Puteneers combined Houdini with 3D GANs as a form-finding tool to test the algorithms' capabilities in creating artifacts as AI agency
~\footnote{Source: \url{https://putteneersjoris.xyz/projects/Ugly\%20Stupid\%20Honest/ugly_stupid_honest.html}}.  
Similarly, Cakmak added a pair of encoder-decoders to a GAN framework to process and save data for generating 3D models~\cite{ccakmak2022extending}. 
This approach aims to extend design cognition by incorporating AI as an agent to understand the mind in cognitive science through embodied action.

Researchers have also increased their focus on training and deployment techniques for design purposes. For instance, Mueller developed a new GAN architecture capable of generating building geometry with incorporated structural features in the dataset~\cite{mueller_3d_2023}. 
This work emphasizes testing different inputs for the generator and discriminator, analyzing the training process, and exploring various hyperparameters. Such progress demonstrates a commitment to autonomously generating building geometry using 3D GANs, leveraging automation and advanced technologies to address architectural and urban design challenges.

Although GANs are widely recognized as the primary tools for deep-learning-assisted architectural generation, they exhibit limitations in architectural design applications, including issues with style variation, category singleness, design unpredictability, and topological inconsistency. These challenges drive designers and researchers to explore diverse alternative approaches in architectural design.

\subsection{Approaches Using Latent Space of VAEs \label{3DVAE}}
VAEs result in an accurate 3D generation but lower resolution, compared to GANs' uncertain while high-quality output. Thus, their applications are commonly generated for different criteria and scenarios against visual output. For instance, Azizi et al. generated reliable and plausible architectural compositions by a VAE through encoding and decoding information about people's spatial movements and activities~\cite{azizi2020floorplan}.


\subsubsection{3D Shape Generation with VAEs}
VAEs utilize pointwise loss to find a probability density by explicit representations to obtain an optimal solution by minimizing a lower bound on the log-likelihood function. 
Brock et al. pioneered the first VAE for processing 3D voxel grids to address the instability issues inherent in GAN approaches~\cite{brock2016generative}. This method employs a pair of encoder-decoder architecture: the encoder, comprising four 3D convolutional layers, maps information to latent vectors, while the decoder reconstructs these vectors into 3D voxels. Subsequent work has focused on enhancing the quality of voxel reconstructions, particularly in producing smooth rounded edges \cite{mittal2022autosdf}.
For point cloud representation, the challenges associated with GANs' instability have prompted the development of alternative generative models based on encoders. Variational autoencoders (VAEs) and adversarial autoencoder models (AAEs) have emerged as viable alternatives \cite{zamorski2020adversarialAE}.
Similarly, generating meshes with VAEs presents challenges akin to those faced with GANs. Given the complexity of processing topology, approaches like the multi-chart parameterization of meshes have been proposed to handle irregular mesh structures \cite{ben2018multi}. Several methods have sought to simplify this process, including SDM-Net-based approaches like TM-Net, which defines a textured space on a template cube mesh \cite{gao2019sdm, gao2021tm, nash2020polygen}.

\subsubsection{VAEs Utilized in Architecture}
For research on architectural generation, VAEs extract information through the latent space with a pair of encoder-decoders, which is compatible with various design purposes but is seldom utilized in architectural design due to high technical access. 
As one pioneering research of 3D generation, Miguel et al. used labeled connectivity vectors extracted from rectangular cube structure "3D-canvas" as data representation because of the usability and compact regarding 3D feature structure~\cite{miguel_deep_2019}. 
The outcome is achieved by 3D voxelized wireframes of multiple architectural forms with different styles and strengths through 
learning continuous latent distributions of VAE. Based on this work, Miguel et al. further tested parametric augmentation for a larger dataset of 3D geometries~\cite{Rodríguez2020}.
By contrast, some work concentrates on post-integration assisting in training according to the designer's preference after the VAE generation in the whole production. 
For instance, Kahraman et al. produced a series of chairs with 3D voxelization while comparing multi-object VAE and GAN~\cite{kahraman_augmenting_2021}. This application discussed pre-defined criteria ranging from leisure to work scenarios. 

Although VAEs are highly capable of incorporating 3D features, their current application is limited to generating specific targets with similar features, such as single buildings. They also produce low-resolution outputs because they use a variational loss to approximate the probability density by optimizing a lower bound on the log-likelihood function. Additionally, the high technical threshold hinders research and applications for non-experts in AI.

\subsection{Approaches Using 3D-Aware Image Synthesis \label{3DAware}}

Most 3D-aware image synthesis methods utilize GAN-based models to sample latent vectors and decode them into target 3D representations. Thus, they offer 3D view-consistent rendering, efficient representation, and interactive editability, as well as integration capabilities with other deep learning generation techniques \cite{surveyshi2022deep}, which could contribute to virtual architectural generation.


  \subsubsection{3D-Aware Image Synthesis and Its Editability}

3D-aware image synthesis achieves 3D view-consistent rendering by relying solely on supervised 2D images and employing different neural rendering techniques. 
Recent research has focused on integrating GAN-based models for generating 3D-aware images \cite{nguyen2019hologan, chan2021pi}. For instance, HoloGAN is capable of unsupervised learning from unlabeled 2D images, eliminating the need for pose labeling, 3D shape, or consistent views \cite{nguyen2019hologan}. It represents a significant advancement in unsupervised learning from natural images. 
Studies, such as Pi-GAN~\cite{chan2021pi}, StyleSDF~\cite{or2022stylesdf}, and StyleNERF~\cite{gu2021stylenerf}, have demonstrated improvements in two critical areas for 3D-aware synthesis: resolution and multi-view consistency of synthetic images. For instance, StyleSDF, based on signed distance functions (SDF), produces detailed 3D surfaces, yielding visually and geometrically superior results \cite{or2022stylesdf}.
Furthermore, cutting-edge techniques integrate 3D-aware images with CLIP models, enabling 3D geometry generation from natural language descriptions \cite{jain2022zerodreamfields, wang2022clipnerf}. DreamFusion, for example, enhances generation efficiency through a loss derived from the distillation of a 2D diffusion model \cite{poole2022dreamfusion}. 
Notably, advancements in 3D-aware synthesis extend beyond pre-trained 2D image-to-text models, with methods like DreamFusion incorporating diffusion models as strong image priors, further improving generation efficiency. Additionally, 3D-aware synthesis can integrate other deep generative models, such as recent diffusion models, which will be discussed in the following subsection.

The primary objective of 3D-aware image synthesis is to achieve explicit control over camera pose, enabling more engaging and interactive user interactions with scenes. Some approaches also support object pose editability; for example, GIRAFFE allows panning and rotating 3D objects in the scene \cite{niemeyer2021giraffe}. Similarly, StyleNeRF enables alteration of style attributes, supporting style blending, inversion, and semantic editing of the generated results \cite{gu2021stylenerf}. This editability offers various solutions for refining generated targets from different perspectives.

\subsubsection{3D-Aware Image Utilized in Architecture}
3D-aware synthesis shows promise for virtual architecture regarding representation conversion, controllability, and multi-modality with linguistic descriptions. 
As OrEl et al. demonstrated in StyleSDF, converting implicit representations to explicit geometry enables designers to comprehend and edit architectural elements with precision visually~\cite{or2022stylesdf}. Similarly, Gu et al. showcased in StyleNeRF how explicit control over style attributes allows for creating diverse architectural designs with semantic editing capabilities~\cite{gu2021stylenerf}.
Recent advancements include Meng et al.'s introduction of a Colab configuration for creating buildings based on DreamField, which supports conditional text input, parameter adjustment, and style attributes~\footnote{Source:~\url{https://github.com/shengyu-meng/dreamfields-3D}}. Additionally, Guida et al. have generated buildings with 3D representation by leveraging NeRF from generative images produced by Stable Diffusion~\cite{guida_gaining_nodate}. This method expands the potential for generating a wider variety of complex forms. 
This implementation underscores the practical application of 3D-aware synthesis in architectural design, providing designers with intuitive tools to realize their creative vision. 
However, concerns arise regarding its effectiveness, which is often limited to simple objects lacking high resolution and precise internal structures due to non-design datasets. 
To address these limitations, Guida et al. utilized integrated stable diffusion with text-to-image synthesis to supplement non-design datasets, effectively enhancing the diversity of form-finding~\cite{guida_gaining_nodate}. 
Nevertheless, the efficiency of generating complex architectural structures with both internal and external details remains constrained by the characteristics of deep generative models.

\subsection{Emerging Approaches Using Diffusion Models \label{3DDPMs}}
3D diffusion models are capable of high quality with fine details in a robust and controlled way for 3D generation, particularly complex shapes with intricate details and specific attributes. 
Thus, their controllability and descriptive input will contribute to deep-learning-assisted virtual architectural generation in the future.

\subsubsection{3D Diffusion and Its Manipulability}
Recently, 3D diffusion has gained a growing interest in generating 3D shapes due to its outperformance abovementioned. 
Despite the flexibility of conditional diffusion sampling, previous diffusion models are limited to sampling pixels of images. 
In contrast, Ben et al. have innovatively applied diffusion models to generate 3D models directly through applying denoising in high-quality 3D-aware image synthesis named DreamFusion~\cite{poole2022dreamfusion}. 
This work sparkes various follow-up approaches regarding text-to-3D (e.g., Magic3D~\cite{lin2022magic3d}, ProlificDreamer~\cite{wang_prolificdreamer_2023}) and image-to-3D (e.g., PVD~\cite{zhou20213dPVD}, Magic123~\cite{qian_magic123_2023}, MVDream~\cite{shi2024mvdream}) with improved generation efficiency, resolution of geometry and texture.
For instance, LION concentrates on greater flexibility regarding operation and application by leveraging conditional synthesis and shape interpolation within a combination of diffusion models and VAE~\cite{zeng2022lion}. 
While Magic123 with image input focuses on reconstructing a high-resolution and textured 3D mesh using joint 2D and 3D priors~\cite{qian_magic123_2023}.

Diffusion models facilitate the generation of 3D shapes with outperformed controllability in two ways: modifying style attributes and manipulating conditioned descriptive text. Firstly, most methods allow modification of specific style attributes, such as shape and texture, during the generation process~\cite{gu2021stylenerf, niemeyer2021giraffe, jain2022zerodreamfields}. This progress suggests that diffusion models offer a robust and controlled approach to 3D shape generation, particularly for complex shapes with intricate details and specific attributes.
Secondly, some burgeoning methods feature text-to-3D generation with human interpretative intention by manipulating text prompts~\cite{chen2018text2shape, poole2022dreamfusion, liu2022towards, lin2022magic3d, nichol2022pointe}. 
Earlier work utilized CLIP for 3D generation and manipulated text prompts, such as PointE~\cite{nichol2022pointe}. Comparably, current works have improved output results by incorporating stronger synthesis priors, such as DreamFusion~\cite{poole2022dreamfusion}.

\sethlcolor{yellow}
\subsubsection{3D Diffusion Model Utilized in Architecture}
Pioneering design works utilizing diffusion models have proliferated recently. While most of these works only utilize diffusion models to generate 2D images coupled with additional computational methods for 3D transposition, such as LoRA and NeRF~\cite{guida_gaining_nodate}.
For instance, AIG studio's work~\footnote{Source:~\url{https://www.bilibili.com/video/BV1Qb411Z7UP/}} has achieved
a modeling framework through stable diffusion, generating plan drawings by relying on heatmaps of urban morphology and additional LoRA models.
Comparably, 
an alternative method employs diffusion models for nonlinear shape interpolation to create a variety of 3D forms~\cite{koh_ai-bewitched_2023,liu_diffusion_2024,sebestyen2023gaining}. For example, one investigation utilized a 3D point cloud probabilistic diffusion model to generate Taihu stone in specified target shapes~\cite{liu_diffusion_2024}. Similarly, another study adopted volumetric density grids as the 3D representation within denoising diffusion models, providing a range of possibilities for architectural design~\cite{sebestyen2023gaining}.

Some initial research in design has also leveraged diffusion models regarding integrating with human interpretation, interaction usability, and multimodality.
An earlier work by Zhang developed a framework capable of compiling analytic semantic information from input texts, then generating 3D architectural form according to descriptive language~\cite{zhang_text--form_2020}.
In this framework, different usages of spaces were trained with adjacent matrices to understand the linguistic instructions. 
The approaches of modifying attributions and conditioned descriptive text broaden the usage case for other layman users by interpreting textual prompts. For instance, Magic 3D has developed a toolkit that offers advanced control over image instances of 3D generation regarding stylization and content based on text prompts~\cite{lin2022magic3d}.

Additionally, recent work has contributed to integrating tools to improve common tedious workflows~\cite{guida_multimodal_2023}, such as constantly transforming models between multiple software or methods.
In this work, Guida developed a plug-in with a user interface embedded in Rhino, integrating multiple approaches of 3D form generation for designers~\cite{guida_multimodal_2023}. This work underscores the significance of multimodality for designer demands. 
The cutting-edge diffusion model's high resolution, denoising, and rapid generation democratizes 3D geometry generation, providing access for individuals with varying levels of expertise to produce creatively. 


\section{Characteristics of Generated Virtual Architecture in Designer-AI Collaboration\label{sec:chara}}

We found some shared methodology characteristics of generative building utilizing deep learning, including scale, scope, operability, and autonomy. Each characteristic has a different degree regarding different design goals and generative models.


\subsection{Scale}
Scale refers to the incorporated model sizes regarding generative capability in architecture, encompassing both the scales of the models themselves and the design context. Large scale pertains to generated city and street areas with diverse design contexts, such as landscapes and river areas or functional planning. Conversely, small scale refers to a generated single building with exterior or interior concerns while lacking environmental design context.
The degree of scale is influenced by two factors: the capability of incorporated composite workflows and the divergent features or capabilities of generative models. 

Divergent generative frameworks possess different capabilities in generating model scale. Firstly, the capability of incorporated composite workflows enables the generation of larger-scale models in multiple phases or modules in a generation. For example, Tencent proposed a large-scale automatic generation solution of 3D virtual scenes incorporating various methods in three modules: city layout generation, building exterior generation, and interior mapping generation \cite{tencentailab_2023}. GANs exhibit such compound capability. Secondly, visual processing capabilities in deep generative models like GANs facilitate the generation of larger-scale models utilizing plan images. For instance, Kim et al. adopted two subtasks of GAN and CNN (convolutional neural network) to construct a 3D city model through scene parsing, city property vectors, and terrain maps. In this work, terrain maps of the city plans are crucial for generating a large city model from the plan.

Furthermore, different models possess various tolerances regarding design context. Firstly, a more hybrid workflow leads to a wider range of design contexts. For instance, the diffusion model can handle broader design contexts since it can merge mixed methods for generation, such as vertical generation by depth map, 3D synthesis by NeRF, and descriptive language. Secondly, the higher capability of accommodating data diversity and complexity results in a wider range of design contexts, such as with GANs and diffusion models. In contrast, VAEs, while proficient in learning latent representations and generating data, are limited in capturing complex spatial and contextual relationships due to model complexity and technology limitations. For example, Miguel’s VAE model to generate a single wireframe 3D building worked with around 150 million parameters, accompanied by only 10\% of the dataset's samples \cite{miguel_deep_2019}. Therefore, working with larger geometries requires a significantly increased amount of data to balance training and validation. Additionally, the number of input parameters scales proportionally with the bounding volume of the input geometry when scaling to large structures, including conditioned context, increasing exponentially.

    \subsection{Scope} 
Scope refers to the range of generation or collaboration involved in a design process with deep learning assistance.
A wide scope typically involves employing more complex processing techniques to delve into deeper relationships between design and generative modeling. Conversely, a specific scope usually entails addressing a particular and tangible objective. Recent work has concentrated on complex processing with a wide scope, such as semantic interpretation \cite{jeffrey_huang_gans_2021, sebestyen_towards_2021}, and latent space understanding \cite{huang_visible_nodate}, in addition to incorporating additional workflows for design criteria \cite{kahraman_augmenting_2021}, particularly in 3D solid form generation approaches. This emphasis stems from the fact that design fields are in the early stages of development and require a comprehensive understanding of deep learning.

The degree of scope depends on two aspects: the involvement of process scope in the design and deep learning-assisted design or human-assisted design. Firstly, involvement in more design phases represents a wider scope, including data training and processing, solution set generation, and evaluation (as illustrated in Section~\ref{ad-ml}). For instance, Mueller conducted the entire scope of design work involving pre-processing datasets, refining and fine-tuning models, and evaluation based on GANs to test the impact of deep learning on design output \cite{mueller_3d_2023}. By comparison, Zhang et al. merely applied GANs to generate a set of section images for 3D transposition without involving other phases, representing a specific scope~\cite{zhang20193d}. 

Secondly, the work involving human-assisted design has a wider scope than deep learning-assisted design regarding design generation since it accommodates additional processes beyond design generation, such as dataset augmentation~\cite{miguel_deep_2019} and semantic labeling~\cite{zhang_text--form_2020}. For instance, Huang et al. extended architects' capabilities by combining the GAN generation with analytical interpretation processes of NLP (natural language processing) instead of solely focusing on design generation~\cite{jeffrey_huang_gans_2021}. Such research addresses the limitations of DL-assisted design generation and compensates for them by incorporating a wider scope of design perspectives. In contrast, most studies relying on generative images only investigate DL-assisted design without addressing other concerns~\cite{zhang20193d,yu2020reprogramming}.

    
\subsection{Operability}
Operability refers to the extent of technical or design operation involved in design collaboration with DGMs. Low operability indicates the straightforward application of DGMs to design tasks, while high operability signifies the ability to operate within the technical space in a work with improved functionality and efficiency. Technical operations encompass various types across different models, including fine-tuning model parameters \cite{mueller_3d_2023}, refining training procedures \cite{ccakmak2022extending, mueller_3d_2023, sebestyen_towards_2021}, customizing architecture \cite{kahraman_augmenting_2021}, and adapting methodology applicability \cite{huang_visible_nodate}.

The degree of operability is influenced by the complexity of technical operations and the applicability of methodology. 
Firstly, the different focuses of design work contribute to varying levels of technical operations and the applicability of the methodology. For instance, Veselý demonstrated high operability by unifying methodology applicability and fine-tuning GAN models for different scales and scopes, including adjusting hyperparameters and optimizing the training process to achieve better convergence and output quality \cite{ondrej_vesely_building_2022}. 
Similarly, Huang et al. explored the latent space potential by connecting structured segmentation of text and visual outputs in generative images of the diffusion model, establishing robust analytical frameworks operation \cite{huang_visible_nodate}. 
Secondly, different designs exhibit varying levels of applicability due to model divergence. Generally, VAEs require relatively high operability from professionals since it processes low-dimensional 3D feature data of buildings. In contrast, diffusion and GAN offer various levels of operability with flexibility ranging from application to fine-tuning, while 3D-aware synthesis requires an intermediate level of operability.

\subsection{Autonomy} 
Autonomy refers to the level of automation in the generation process facilitated by deep learning techniques. High autonomy indicates a largely automated process driven by a unified workflow, while low autonomy involves manual intervention, requiring the connection of workflows and greater ease of use.
In architectural design, most works exhibit relatively low autonomy to ensure quality and design accuracy under the guidance of designers. For example, Mueller validated the methodology across diverse scenarios by selectively retrieving spatial data and comparing the performance of proposed generative models \cite{mueller_3d_2023}. However, recent advancements show a proliferation of high autonomy, especially in 3D generation, focusing on task efficiency~\cite{kim_citycraft_2020, tencentailab_2023}. 
Nevertheless, there is a growing trend towards higher autonomy in design work to broaden access and enhance usability. For instance, Guida developed an integrated tool within the visual scripts of Grasshopper with high autonomy, combining all processes and modules in architectural design, providing convenient input and output modules across different approaches \cite{guida_multimodal_2023}.

The degree of autonomy is influenced by the complexity of the design purpose and model processing requirements. Firstly, researchers tackling challenging and complex design objectives, such as connecting generative models with design output, tend to exhibit low autonomy. For instance, Mueller endeavored to design goals and each step of generative models, including training datasets and key hyperparameters~\cite{mueller_3d_2023}. 
Secondly, the model processing with frequent transposition across platforms and tools in design leads to low autonomy. For example, Özel incorporates additional procedural steps after generating a 2D image, requiring high reliance on manual designer intervention~\cite{ozel2020interdisciplinary}. 
Thirdly, the absence of unified criteria further leads to low autonomy. Özel's work, for instance, involves experimenting with various 3D modeling software and photogrammetry, indicating low autonomy and hindering further development in architectural design towards~\cite{ozel2020interdisciplinary}.

\section{Research Agendas\label{sec:researchagendas}}
\begin{figure}
    \centering
    \includegraphics[width=0.99\textwidth]{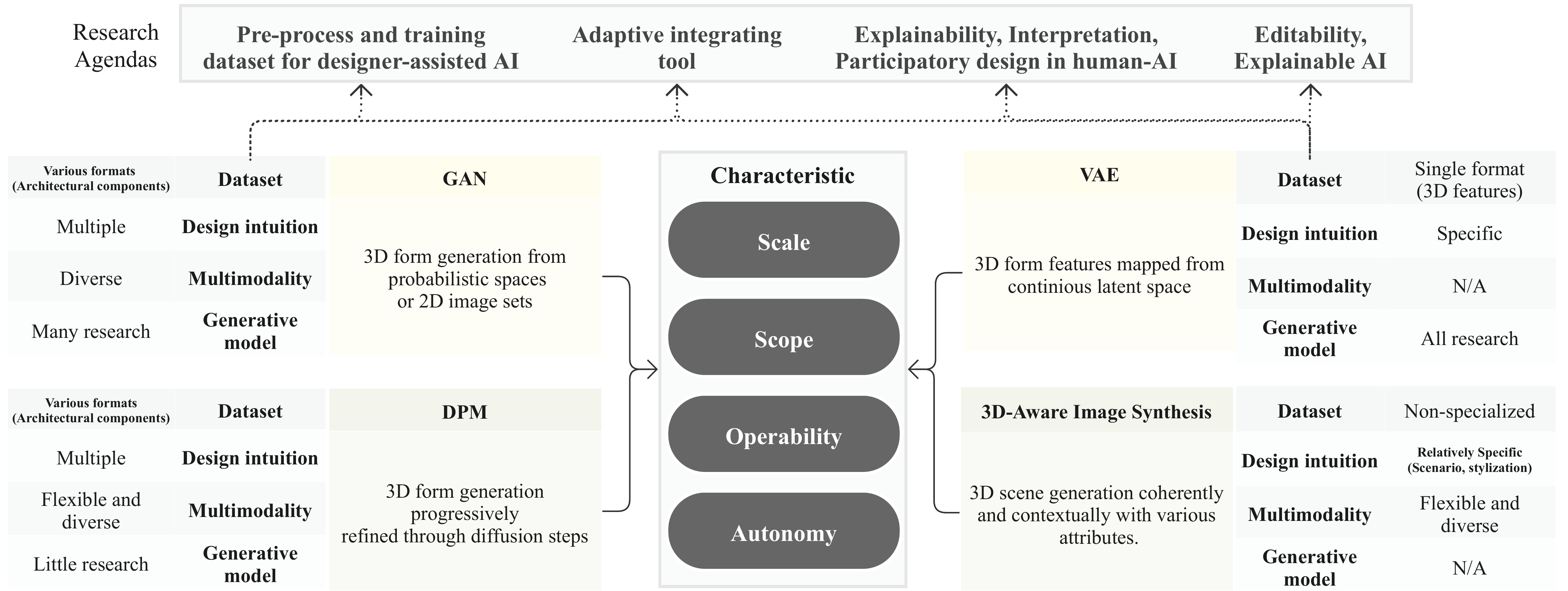}
    \caption{A summary diagram illustrating the connections among key sections, including the overview (Section~\ref{sec:overview}), generative approaches (Section~\ref{sec:DGMs}), and characteristics of these approaches (Section~\ref{sec:chara}). Accordingly, these high-level issues, identified within the four key sections of our survey, serve as the basis for further research agendas (Section~\ref{sec:researchagendas}).}
    \label{fig:researchagenda}
\end{figure}

In the previous sections, we explored the urgent need for HCI and DL-assisted methods in generated virtual architecture. We thoroughly examine the literature and elucidate four key focuses, as mentioned in Section~\ref{ad-ml}. These key focuses serve as the foundation for our research agendas, formulated from a perspective of high-level questions (Fig.~\ref{fig:researchagenda}). Our investigation of current literature and works also reflects on these questions. Therefore, we advocate the following research agendas to promote research in this area.


\subsection{Agency Between Designer and Machine in Data Processing and Training} 
Agency between humans and machines emerges as a significant feature in AI-assisted design and designer-AI collaboration. On the one hand, a human-driven agency faces challenges due to low-resolution and incomplete datasets, requiring human intervention \cite{shi2023understanding}. On the other hand, an AI-driven agency raises concerns about intellectual property issues 
but offering higher efficiency~\cite{guida_multimodal_2023}. Our analysis of the survey identifies two main barriers to achieving a balance between human and AI agencies: flexible datasets and human-assisted machine learning.


Flexible and comprehensive datasets are crucial for achieving balanced agency, supporting various design purposes with precision and semantic information. However, limitations in dataset availability pose significant challenges in both industries and academia. As noted by Regenwetter et al., three main issues result in this challenge: lack of available 3D datasets, insufficient data size, and data sparsity and bias \cite{regenwetter2022engineering}. 
 The limited availability of 3D building datasets restricts massive computation in 3D solid generation, hindering designers' ability to find appropriate datasets.
Correspondingly, our review expounds that only a few studies contributed to the 3D building dataset (N=3). 
While datasets like Point-E and GET3D show promise in improving efficiency and reducing designer labor, architectural work highlights the complexity of preprocessing datasets to suit various building scales and design purposes.


In addition to addressing dataset limitations, human-assisted deep learning could be a viable solution in design work, including training (e.g., semantic analysis~\cite{zhang_text--form_2020, sebestyen_towards_2021, jeffrey_huang_gans_2021}) and regulation (e.g., post-processing tasks~\cite{kahraman_augmenting_2021}). 
For example, Liu et al. regulated AI's behavior for screening target outputs with manual labels based on certain architectural criteria of 3D form configurations from a generative script~\cite{liu_anonymous_2019}.
Our analysis of the survey reveals that human-assisted machine learning is an emerging and underexplored method,  with few studies focusing on this area (N=4). It holds significant potential for specialized design and production pipelines in DL-generated architecture, such as building semantic interpretation and regulating design criteria. Accordingly, 
some researchers question whether future designers working with training data in established deep learning workflows may spend more time than on result output~\cite{sebestyen_towards_2021}. 
Additionally, data customization could offer a solution by tailoring datasets to specific formats, precision, and sizes for multiple design purposes, requiring collaboration among enterprises, managers, platforms, and users.
Therefore, further research on agency between generation and design requires close collaboration with technicians, designers, and producers through human-assisted machine learning.

\subsection{Augmenting Communication Between Design and Artificial Intelligence} 
To benefit the design project's fitting purpose, HCI researchers and designers should augment communication with each other by promoting tools or enhancing workflows between design applications and deep learning. We found two aspects closely related to this agenda: explainability in designer-AI collaboration and editability in the generative model.

        \textit{Explainability.}
Explainability is one recognized principle of communication within designer-AI collaboration, essential for fostering trust in AI systems and guiding the design process and outcomes~\cite{mohseni2021multidisciplinary}. However, a significant disparity exists in the developmental progress between HCI and the design field. For instance, HCI researchers have made tremendous strides in responsible AI for explainability, focusing on user experience, trustworthiness, and collaboration~\cite{SAEED2023110273}. The design field lags due to existing challenges and a lack of expertise among designers in negotiating input, process, and controlling output. According to our analysis of the survey, no one has addressed this critical issue. Regardless, recent studies (N=2) have primarily investigated methods for controlling output by manipulating detail in algorithms within generative models to achieve greater controllability in design outputs. 
Therefore, future research in HCI and design should delve deeper into explainability in design and generation processes building upon design principles.

        \textit{Editability.}
Editability is the other key to communication, relying on timely feedback and adjustments to model 3D objects. Similarly, there is a big gap in development progress between the generation technique and design fields.
 Prior work has demonstrated editing autonomy in 3D shape generation techniques available to users. For example, Liu et al. proposed a user-centric 3D shape generative method by drawing a target model as 3D voxel grids~\cite{Liu2017Interactive3M}. While a few works demonstrate some indirect approaches to editing 3D models with implicit representations \cite{hao2020dualsdf, ibing20213d, deng2021deformed, zheng2021deep}. According to our survey, although none of the architectural works mainly contribute to editing in 3D generation, one indicated a potential for variability in 3D form. This work generates infinite continuous forms by interpolating 3D features among multiple wireframe representations with compact implicit functions~\cite{miguel_deep_2019}. Through this interpolation with implicit representations, we could envision a broad use for virtual architecture in supporting itself with real-time dynamic morphology. 
 Thus, the editability potential of implicit representation incorporating building form launches the future agenda for deep learning virtual architecture design.

In contrast to 3D shape generation, 3D-aware image synthesis precludes direct user manipulation in 3D space and requires latent vector editing to control the composition, shape, and appearance. These latent vectors model all other variables that are not captured by physical factors while being able to control small changes in the scene, such as lighting and coloring~\cite{xia20223dawaresurvey}. Furthermore, several approaches allow additional inputs to alter the editing of the scene, such as textual descriptions, semantic labels~\cite{liu2022towards, jahan2021semantics, fu2022shapecrafter}, images~\cite{sanghi2022clipforge}, and parameter controls. 
Therefore, this approach could be the complementary editing layer to deep learning-assisted design. For instance, 3D representation from a user end in a UI panel can be controlled according to architecture settings. Therefore, this editability is one of the key agendas for the rapid popularity of virtual architecture with AIGC-driven 3D designs.

\subsection{Enhancing User Consideration with Diversity and Collectives under Designer-AI Collaboration}
Designs involving multiple users necessitate inclusivity by incorporating usage affordances and considering wide needs for virtual architecture since the virtual environment pertains to shared environments for collectives and individuals from diverse backgrounds and cultures through converging and engaging, leading to highly multi-social experiences~\cite{jonassen1999activity, bartle2004designingvw}. 
Accordingly, we identify two focal conflicts regarding this agenda: the balance between collective and individual, as well as culture and efficiency.

\textit{Efficiency and Diversity.}
Deep learning enables the rapid creation of large-scale architectural environments but often lacks diversity in catering to multiple users and their objectives. Conversely, diverse buildings may exhibit non-uniform quality criteria, such as inconsistencies between manually and automatically generated structures~\cite{keil2021creating, yao20183dcitydb}. Regarding this, spatial accuracy and level of detail are significant criteria for such virtual building quality~\cite{yao20183dcitydb}, including scale, size, granularity, and simulation fidelity ~\cite{dionisio20133dmetaverse}.
Additionally, diversity also leads to a high cost of human resources~\cite{yao20183dcitydb}. 

Reliable solutions entail unified pipelines and highly automated computing or provide user-friendly toolkits with standardized governance by platforms. For example, Tencent AI Lab proposed a high-automated solution for generating 3D virtual scenes with a unified pipeline consisting of various modules, supporting layman with industry production~\cite{tencentailab_2023}~\footnote{Source:~\url{https://gdcvault.com/play/1028921/Recorded-AI-Enhanced-Procedural-City}}. Similarly, Mozilla Hubs' Spoke offers a robust toolkit for 3D scene editing catering to a wide range of users.

\textit{Individuals and Collectives.}
A significant challenge arises from the conflict between individual-centric ideologies of new liberalism and the collective nature of the virtual environment~\cite{nazmeeva2019constructing}. While the exhibiting of ownership, identity, and recommended content prominence the precise value of an individual in expansive virtual spaces under new liberalism~\cite{nazmeeva2019constructing}.  Within this dynamic environment, we are increasingly accused of proliferating products manifesting our unique identities and personal needs~\cite{nazmeeva2019constructing}. This situation leads to an urgent rethinking of the collaboration with virtual worlds regarding the relationship between collectivism and individualism.

One potential solution is public participation in the design process. Designers have conducted participatory design experiments positively, addressing multifaceted problems and designing for minority groups. 
Participatory design allows non-professionals to make design decisions collaboratively alongside professional designers.
The emergence of creative collectives in architecture underscores the importance of trends in public participation~\cite{eloy_virtual_2021}. However, our analysis of the survey found no research on participatory concerns for 3D generation or DL-assisted architectural design. In contrast, there has been a significant achievement in participation topics within designer-AI collaboration~\cite{simonsen2020infrastructuring}. Therefore, we urge designers and researchers to emphasize the public participatory potential for virtual building by referencing existing research.

\subsection{Designing Adaptive Tool by Integrating Multiple Tools with Multimodal User Input and Output}

Integrating tools results in wide access with a user-friendly interface towards rich user-created content in the future metaverse. 
Although this vision has been underscored in the virtual environment industry, the adaptive tool for diverse users to design fields is unseasoned. The reason behind this agenda is the lack of applications that integrate multimodality and consider user experience.

\textit{Multimodality.}
As aforementioned, the previous tremendous progress in DL-assisted architectural design has been built upon "visual" representation with the generative image since "visual" modality as design essential directly links design purpose and output representation~\cite{shi2023understanding}. Natural language has also been pivotal in multimodality for design. As Markus emphasized, "language is at the core of making, using, and understanding buildings" ~\cite{forty2000words}. Correspondingly, our analysis of the survey underpins future trends from the latest boom in text-to-3D and 2D-to-3D approaches in the 3D generation field. However, very few architectural designs concentrate on this method due to its novelty.
Compared to the industry, huge success has been applied to user-end industrial tools, such as Midjouney and DALLE2, which have wide access to diverse designers with various purposes and alter the design workflows. Given this sense, the other modality combined with visuals could boost emerging designs from tool and technical perspectives, especially for virtual architecture from 3D generation. 
Therefore, bridging visuals and language with 3D form as one research agenda could reform architectural design, particularly in a virtual environment.

\textit{Integrating Tool.}
User-friendly tools would also inspire broadened usage, facilitating designers' work.
In this regard, we anchored significant positions to illuminate this agenda from the industry-research gap and the technique-application gap according to the analysis of our surveyed paper.
Firstly, current architectural work still focuses on architectural forms and innovative design methods connecting designers and techniques. According to our analysis of the survey, only one research developed integrated tools for architectural design generation (N=1).
Secondly, most research in CV fields contributes novel algorithms and frameworks, ignoring further utilizing technique innovation in the design application.
Yet integrating tools is crucial to bridge industries, and research facilitates future virtual architecture development. 
Our investigation provides such a critical lens for understanding the gap between architecture and such models to further utilize such generative models for virtual architectural design, including Dreamfusion3D~\cite{poole2022dreamfusion}, Point-E~\cite{nichol2022pointe}, and Get3D~\cite{gao2022get3d}.
Therefore, we appeal to interdisciplinary research to conduct these untilled topics to bridge the gap between the two fields to follow the industry's advanced progress, especially for HCI and design researchers.

\subsection{Research Directions and Future Opportunities}
Our survey unveils that this field necessitates interdisciplinary collaboration at the intersection of 3D generation, virtual reality, sensing technologies, and wearable computing. 
Thus, we address three future directions of generated virtual architecture.

\subsubsection{Designing User-Centric Adaptive Virtual Buildings}
Recent advancements in wearable technology incorporating sensors and actuators have significantly enhanced the study of virtual spaces within a design framework~\cite{diniz2019body,BOWER2019101344}.
Virtual architecture research has positively conducted interaction with the virtual environment by collecting user bio-signal data~\cite{narahara_kurashiki_2022,pei_biofeedback_2020, homolja2020impact,ding_spatial_2022,sheehan_fourth_2021,nguyen2019koala}.
This trend embodies that feedback and interaction between the built environment and humans are crucial for creating livable and adaptive virtual environments.
Two significant opportunities for future exploration emerge within this context:
Firstly, integrating emotional computing systems into 3D generation approaches is crucial for designing emotionally resonant spaces~\cite{homolja2020impact}. For example, Sheehan et al. demonstrated the simulation of atmospheric qualities in VR to explore space and directionality, using thermal perception as feedback~\cite{sheehan_fourth_2021}. 
Secondly, utilizing real-time data captured by ubiquitous computing and wearable devices allows for personalized and responsive user interactions in virtual environments~\cite{guo_method_2021, zarei_design_2021}. For instance, Tosello creatively translated physiological data in the presence of virtual scenarios, including thoughts and emotions, into digital space~\cite{tosello2003performing}. This approach facilitates the creation of interactive virtual architectures that dynamically adapt to user behavior and preferences in real-time, thus expanding the scope of research in responsive design. For instance, users could customize virtual buildings in real-time by adjusting parameters and weights in generative models based on ubiquitous data collected from wearable devices.
Therefore, we envision the future of virtual architecture to be centered on interactive, self-responsive forms generated by deep learning approaches. 

\subsubsection{Engaging Users in Governance and Content Generation: A Role Shift as Consumers.}
Virtual buildings, serving as 3D digital resources of content and space, play a pivotal role in fostering the content ecosystem in the virtual environment, which involves consumers, creators, and platforms. 
In a recent study, Ding et al. demonstrated an innovative approach to user-personalized architectural design in virtual space, utilizing a user-friendly tool for designer-AI collaboration~\cite{ding_exploration_2022}. 
 This tool showcases an interface and features tailored for creating virtual architecture, aiming to ensure accessibility and ease of use. 
Tools featuring user-friendly interfaces and customization options can empower consumers to assume governance and content-generation roles.
Therefore, we believe multiple approaches enabling user engagement in various roles represent the prevailing trend within the virtual environment context.




\subsubsection{Generating Innovative Virtual Forms and Dynamic Morphology with Social Sustainability.}
Deep learning is propelling virtual architecture towards innovative forms shaped by algorithmic principles, surpassing conventional visual geometries. 
The algorithmic form encapsulates the relationship between computational processes and the resulting object characteristics~\cite{chaitin1975theory, chaitin1975randomness}. For instance, recent research has proliferated in aesthetics assessment~\cite{tonn2017designing,kavakoglu_computational_2021,sardenberg_aesthetic_2019}, regarding visual character~\cite{stuart-smith_visual_2022}, and visual sense~\cite{wells_beauty_2021}.
In this context, an increasing number of social algorithm proposals integrate social factors like environmental and social sustainability into deep learning frameworks.
For instance, Koh et al. employed 3D GANs to design a series of housing buildings ensuring sustainable economic balance and social resident privacy~\cite{koh20223dganhousing}.
Moreover, the instantaneous feedback loop in environmentally intelligent virtual architecture aids its management and governance, streamlining activities like building reviews and further adaptation. 
This responsive capability fosters a collaborative environment.

\subsubsection{Ethical implications of AI-generated architecture}


In an era of increased collaboration with AI, it is essential to examine the ethical implications of originality and authorship, the morality of machine creation, and human rights within a pervasive data environment. In this context, user privacy and authorship are of utmost importance.

Designers must work closely with AI experts and developers to ensure data privacy, including confidentiality, integrity, and availability. As designers become more involved in AI procedures such as deployment, training, and decision-making as the party of developers, they have a responsibility to raise awareness and take action to protect data privacy~\cite{golda_2024_privacysurvey}.

AI-generated content in virtual digital environments presents significant challenges related to authorship, ownership, and copyright within a complex content ecosystem due to zero-cost reproduction and machine intelligence~\cite{vyas_2022_ethical}. 
For example, original creations can be utilized by AI models to produce suspected pirated content without informing the authors. While the industry offers solutions such as uploading creations with copyright-level options, 
challenges persist regarding evaluation and feasibility, particularly in specific areas~\cite{vyas_2022_ethical}. 
Specifically, the reuse of creations should involve informing the author and obtaining permission. The benefits of distribution or authorship should be considered, especially when users may serve in multiple roles, including potential creators, re-creators, distributors, and regulators. Therefore, designers with a robust understanding of the field and a commitment to authorship play a crucial role in establishing specific criteria to regulate the generation approaches of AI-generated architecture.


\section{Conclusion}
In this survey, we explore various approaches deep neural networks utilize to generate virtual architectures autonomously. Our investigation spans three main domains: architectural design, 3D generation techniques, and virtual environments for designer-AI collaboration. Through an in-depth literature analysis, we focus on four key aspects: datasets, multimodality, design intuition, and generative frameworks. We aim to bridge the current research gap from an interdisciplinary perspective.

Our survey highlights the lack of systematic research, particularly concerning the virtual dimensionality of architecture and the collaborative design aspects of human-AI interaction. We advocate for comprehensive and interdisciplinary research efforts encompassing various themes such as agency, communication, user considerations, tool integration, and the involvement of diverse stakeholders, including technical developers, HCI researchers, architects, and virtual environment practitioners.



\bibliographystyle{ACM-Reference-Format}
\bibliography{sample-acmsmall}

\clearpage
\appendix
\section*{Appendix}
\section{Supplementary Material (e-pub only)}
This material is supplementary and should be included in the electronic version only.

\begin{table}[ht]
    \small
    \centering
    \caption{Overview of the mission and scope of virtual architecture, focusing on five key aspects: spatial forms, user presence, socialization, interaction, and interoperability and scalability.}
    \begin{tabular}{p{1.6cm}p{13.9cm}}
        \toprule
        \textbf{Discipline} & \textbf{Description} \\
        \toprule
        
        Spatial forms & Incorporating diverse forms in terms of function, style, and purpose to ensure efficient production to achieve large-scale outcomes. \\
        \hline
        User presence, experience & Ensuring self-presence with a pleasant, understandable user experience, considering human perception (e.g., comfort, safety), behavioral demands (e.g., navigation, gathering activities), and environmental factors (e.g., lighting, sounds, atmosphere) in immersive environments. \\
        \hline
        Socialization, engagement & Supporting diverse and multiple social interactions, including activities, engagements, and goals for live, entertainment, and work. Providing collaboration opportunities with integrated tools for editing, altering, and customizing tasks. \\
        \hline
        Interactive elements & Designing flexible, dynamic, and real-time interactive human-centric buildings by utilizing ubiquitous data from environmental intelligence, such as responsive or interactive building elements. \\
        \hline
        Interoperability, scalability & Building interoperability and scalability across multiple platforms focusing on data security, user privacy, and resource management for sustainability. \\
        \bottomrule
    \end{tabular}
\label{tab:disciplines}
\end{table}

\begin{table}[ht]
  \fontsize{6pt}{6pt}\selectfont
  \centering
    \caption{This article elucidates four key focuses derived from the research questions reviewed for deep learning-assisted architectural design papers: datasets, multimodality, design intuition, and generative models (depicted as four columns on the right). A selection of the reviewed papers is provided in the table as a reference.}
  \renewcommand{\arraystretch}{1.2}
  \begin{tabular}{
  p{0.05\textwidth}
  p{0.075\textwidth}
  p{0.3\textwidth}
  p{0.14\textwidth}
  p{0.14\textwidth}
  p{0.04\textwidth}
  p{0.04\textwidth}
  p{0.04\textwidth}
  p{0.04\textwidth}}

    \toprule
    
    \multirow{2}{1.5cm}{\textbf{Reference}} &\multirow{2}{1cm}{\textbf{Category}} &\multirow{2}{6cm}{\textbf{Research Question}} &\multirow{2}{1.2cm}{\textbf{3D Representation}} &\multirow{2}{1.2cm}{\textbf{Deep Generative Model}}
    &\multicolumn{2}{c}{\textbf{Key Focus}} \\

    \cline{6-9} & & & & &Dataset & Multi-Modality& Design Intuition&Generative Model\\
    \toprule


 
 
 

 
 




    



\cite{ccakmak2022extending} & 3D Solid & Extend design cognition  & Point cloud/Mesh & GAN (with Autoencoder)&  \checkmark & & \checkmark & \\ \hline

\cite{mueller_3d_2023} & 3D Solid & Generate building massing  & Point cloud/Voxel Grid & GAN & \checkmark  & & \checkmark  & \checkmark \\ \hline

\cite{miguel_deep_2019} & 3D Solid& Generation, manipulation and form finding of structural typologies  & Wireframe & VAE& \checkmark & & & \checkmark \\ \hline
    
\cite{kahraman_augmenting_2021} & 3D Solid & Solve design problems incorporating deep learning & Voxel grid & Based on VAE \& GAN &\checkmark & & \checkmark & \checkmark \\ \hline
    
\cite{sebestyen_towards_2021} & 3D Solid & New way to design models  & Voxel grid & VAE&\checkmark &\checkmark &\checkmark & \\ 

\hline


\cite{guida_multimodal_2023} & Text-3D Solid & Multimodal generation with text input & - & Diffusion model & &\checkmark & \checkmark &  \\ 
\bottomrule

  \end{tabular}
    \label{tab:keyfocus}
\end{table}

\begin{table}[ht]
  \caption{An overview of 3D generative approaches for shape generation, each allowing the creation of editable models for explicit representations.}
  \fontsize{6pt}{6pt}\selectfont
  \centering
  \renewcommand{\arraystretch}{1.2}
  \begin{tabular}{|p{0.22\textwidth}|p{0.22\textwidth}|p{0.22\textwidth}|p{0.22\textwidth}|}
  \hline
  \textbf{Method Names}& \textbf{Publication} \& \textbf{Year} & \textbf{3D Representation} &\textbf{Deep Generative Model}\\
  
    
    \hline
    3D GAN~\cite{wu20163DGANlearning} & NIPS 2016 & Voxel grid & GAN   
    \\ \hline
    Text2Shape~\cite{chen2018text2shape} & ACCV 2018 & Voxel grid & GAN   
    \\ \hline
    PLATONICGAN~\cite{henzler2019escaping} & CVPR 2019 & Voxel grid & GAN 
    \\ \hline
    IG GAN ~\cite{lunz2020inverse} & arXiv 2020 & Voxel grid &GAN  
    \\ \hline 
    
    Achlioptas et al. ~\cite{achlioptas2018learning} & ICML 2018 & Point cloud &GAN  
    \\ \hline 
    Shu et al. ~\cite{shu20193d} & CVPR 2019 & Point cloud &GAN  
    \\ \hline

    Get3d ~\cite{gao2022get3d} & NeurIPS 2022 & Mesh &GAN  
    \\ \hline
    
    IM-Net  ~\cite{Chen_2019_CVPR} &CVPR 2019  &Neural field  &GAN  
    \\ \hline
    Kleineberg et al.  ~\cite{kleineberg2020adversarial} &arXiv 2020  &Neural field  &GAN 
    \\ \hline

    Brock et al.  ~\cite{brock2016generative} &arXiv 2016  & Voxel grid &VAE  
    \\ \hline
    Autosdf  ~\cite{mittal2022autosdf} &CVPR  2022  & Voxel gird  &VAE 
    \\ \hline
    Sagnet  ~\cite{wu2019sagnet} &TOG 2019  & Voxel gird &VAE
    \\ \hline
    Li et al.  ~\cite{li2020learningVAE} &AAAI 2020  & Voxel gird &VAE
    \\ \hline
    Pq-net  ~\cite{wu2020pq} & CVPR 2020  & Voxel gird &VAE 
    \\ \hline
    AdversarialAE ~\cite{zamorski2020adversarialAE} & CVPR 2020  & Voxel gird &VAE 
    \\ \hline
    
    Multi-Chart ~\cite{ben2018multi} & TOG 2018 & Mesh &VAE  
    \\ \hline
    SDM-NET ~\cite{gao2019sdm} & TOG 2019 & Mesh &VAE  
    \\ \hline
    Tm-net ~\cite{gao2021tm} & TOG 2021 & Mesh &VAE 
    \\ \hline
    Polygen ~\cite{nash2020polygen} & ICML 2020 & Mesh &VAE 
    \\ \hline
    
    PointFLow ~\cite{yang2019pointflow} & CVPR 2019 & Point cloud & Normaliz. flow model
    \\ \hline
    
    CLIP-Forge ~\cite{sanghi2022clipforge} & CVPR 2022 & Voxel grid & Normaliz. flow model 
    \\ \hline
    DreamFusion~\cite{poole2022dreamfusion}& arXiv 2022 & Mesh & Diffusion model
    \\ \hline
    PVD ~\cite{zhou20213dPVD} & CVPR 2021 & Hybrid: point cloud-voxel & Diffusion model 
    \\ \hline
    Magic3D ~\cite{lin2022magic3d} & arXiv 2022 & Neural field-Mesh  & Diffusion model
    \\ \hline 
    LION ~\cite{zeng2022lion} & arXiv 2022 & Mesh & Diffusion model
    \\ \hline 
    Point-E ~\cite{nichol2022pointe} & arXiv 2022 & Neural field-Point cloud  & Diffusion model
    \\ \hline 
    Magic123~\cite{qian_magic123_2023} & arXiv 2023 & Mesh & Diffusion model
    \\ \hline
    Prolificdreamer ~\cite{wang_prolificdreamer_2023} & arXiv 2023 & Mesh & Diffusion model
    \\ \hline
    Hifi-123~\cite{yu2024hifi123} & arXiv 2024 & Mesh & Diffusion model
    \\ \hline
    MVDream ~\cite{shi2024mvdream} & arXiv 2024 & Mesh & Diffusion model
    \\ \hline

  \end{tabular}
  
\label{tab:CV3dshape}
\end{table}

\begin{table}[ht]
  \fontsize{6pt}{6pt}\selectfont
  \centering
    \caption{An overview of 3D generative approaches of 3D-aware image synthesis.  `Single/multiple' represents the result generated by a single image adopting a sample of single-view or multiple images adopting multiple-view images. `Geometry' indicates whether this method allows the export to mesh. 
     `Editability' refers to the ability to directly manipulate or modify the synthesized output, such as adding or removing objects or adjusting their properties. `Controllability' refers to the ability to adjust variables during the generation process, such as camera pose, object position, and lighting.
    }
  \label{tab:3Dawareimage}
  \renewcommand{\arraystretch}{1.2}
  \begin{tabular}
    {|c|p{0.1\textwidth}|p{0.1\textwidth}|p{0.07\textwidth}|p{0.065\textwidth}|p{0.065\textwidth}|p{0.065\textwidth}|p{0.065\textwidth}|p{0.1\textwidth}|}
    \hline
    \multirow{2}{1.5cm}{\textbf{Method Names}} &\multirow{2}{1.5cm}{\textbf{Publication \\ \& Year}}  &\multirow{2}{1.5cm}{\textbf{3D Repre-\\sentation}} &\multirow{2}{1.5cm}{\textbf{Single/multi-\\ple-view}} &\multirow{2}{1.5cm}{\textbf{Geometry}} &\multirow{2}{1.5cm}{\textbf{Editability}} &\multicolumn{2}{|p{0.1\textwidth}|}{\textbf{Controllability}} &\multirow{2}{1.5cm}{\textbf{Highlight}}\\
    \cline{7-8}& & & & & &Camera & Object &  \\
    \hline

    NeRF ~\cite{Mildenhall2020NeRFRS} & CVPR 2019 & Neural field & multiple & - &- &\checkmark & \checkmark - & 3D Novel View Synthesis\\ \hline 
    
    HoloGAN ~\cite{nguyen2019hologan} & CVPR 2019 & Voxel grid & single & \checkmark  &- & \checkmark & -  & -  \\ \hline 
    Pi-GAN ~\cite{chan2021pi} & CVPR 2021 & Neural field & single & \checkmark & - & - & - & - \\ \hline 
    Giraffe ~\cite{niemeyer2021giraffe} & CVPR 2021 & Neural field & single&- & \checkmark &\checkmark & \checkmark & - \\ \hline 

    StyleSDF ~\cite{or2022stylesdf} & CVPR 2022 & Neural field & single&\checkmark & - &\checkmark&-&-\\ \hline 
    StyleNeRF ~\cite{gu2021stylenerf} & arXiv 2021 & Neural field &single&\checkmark & - &\checkmark  &-&-\\ \hline 
    
    DreamField ~\cite{jain2022zerodreamfields} & CVPR 2022 & Neural field & multiple &\checkmark& &\checkmark  &-& Text input (CLIP) \\ \hline 
    DreamFusion ~\cite{poole2022dreamfusion} &arXiv 2022& Neural field  & single  &\checkmark  & \checkmark &\checkmark  &\checkmark & Text input \\ \hline 
    CLIP-NeRF ~\cite{wang2022clipnerf} & CVPR 2022 & Neural field &  single &-  &\checkmark &\checkmark  & \checkmark 
    & Text input (CLIP)\\ \hline
  \end{tabular}
\end{table}

\begin{table}
  \fontsize{6pt}{6pt}\selectfont
  \centering
  
  \caption{
  The related works in architecture are classified into three categories: 2D to 3D transposition, 3D solid generation, and text-based 3D form generation. The `Category' column denotes the generation approaches: `2D to 3D' refers to generating 3D forms from 2D images; `3D Solid' involves directly generating 3D forms using DGMs such as GANs, VAEs, 3D-aware image synthesis, and diffusion models; `Text-3D Solid' indicates that the 3D form generation process incorporates text input to control the output. The `Objective' column describes the research directions and goals. The `Methodology' column outlines the proposed workflow for the generation process. `3D Representation' indicates the formats of the generated data, including point cloud, voxel grid, and mesh. `Generative Model' specifies the type of deep generative model used in each work.}

  \renewcommand{\arraystretch}{1.2}
  \begin{tabular}{p{0.05\textwidth}p{0.075\textwidth}p{0.15\textwidth}p{0.4\textwidth}p{0.1\textwidth}p{0.1\textwidth}p{0.1\textwidth}p{0.05\textwidth}p{0.05\textwidth}}

    \toprule
    
    \textbf{Reference}&\textbf{Category}&\textbf{Obejctive}&\textbf{Methodology}&\textbf{3D Representation}&\textbf{Generative Model} \\
    \toprule

\cite{del2019church} & 2D to 3D & Test AI agency in design & Utilizing Style Transfer to train two datasets Baroque and Modern images as a basis to form a 3D model  & - & - \\ \hline

\cite{ozel2020interdisciplinary} & 2D to 3D & HCI in urban design & Utilizing Style Transfer to generate different stylized images and generate 3D geometry through procedural modeling & - & - \\ \hline
 
\cite{ren_spire_2020} & 2D to 3D & Test AI agency in design & Utilizing Style Transfer to replace pixels with voxelization units to generate 3D forms & - & - \\ \hline
 
\cite{liu2021pipes} & 2D to 3D & Toolkits for 3D generation & Utilizing Style Transfer to assist the generation of 3D structure from 2D images & - & - \\ \hline
 
\cite{yu2020reprogramming} & 2D to 3D & Generate building massing  & Utilizing Pix2Pix GAN to generate plan pattern and section pattern, then converted to 3D massing  & - & - \\ \hline

\cite{di_carlo_generating_2022} & 2D to 3D & Generate building massing & Utilizing Pix2Pix GAN to generate urban morphology to create building massing & - & - \\ \hline
 
\cite{zhang20193d} & 2D to 3D & Form finding to assist design & Slicing a 3D model and trained with different combinations of 2D StyleGANnetworks, and finally stitching into a 3D model & - & - \\ \hline
 
\cite{zhang20193d} & 2D to 3D & Form finding to assist design & 3D model generation based on 2D plan and section using Style Transfer  & - & -\\ \hline

\cite{zhang2021machine} & 2D to 3D & Form finding to assist design & Combining the spatial sequence information to generate 3D form from 2D images through multi-level deep generative networks such as StyleGAN& - & -\\ \hline

\cite{bank2022learning} & 2D to 3D & Human and neural network interface & Utilizing 3D solid for training to map spatial semantics to a latent space assembled using point cloud representations & Point cloud & GAN\\ \hline

\cite{asmar2020machinic} & 2D to 3D & Integrate latent space in design & GAN allows for navigation in the latent space to create digital designs using vector arithmetic and interpolation techniques, then converting resulting images to 3D voxel structures & Voxel grid & GAN\\ \hline

\cite{jeffrey_huang_gans_2021} & 2D to 3D & Recognize 2D pattern to 3D form & Utilizing Latent space rotation and perspective projection to generate 3D model & Voxel grid & GAN \\ \hline
    
\cite{kim2022perspectival} & 2D to 3D & Recognize 2D pattern to 3D form & Utilizing Latent space rotation and perspective projection to generate 3D model & Voxel grid & GAN \\ \hline

\cite{ondrej_vesely_building_2022} & 2D to 3D & Generate building massing & Employing voxelization and vectorization techniques on images generated by the Pix2Pix GAN model to generate 3D building massing & Voxel grid & GAN\\  \toprule
    
\cite{ccakmak2022extending} & 3D Solid & Extend design cognition & Utilizing a GAN with a pair of encoder-decoder to process datasets and generate new 3D models, resulting in different 3D representations like point cloud and mesh & Point cloud/Mesh & GAN (with Autoencoder)\\ \hline

\cite{mueller_3d_2023} & 3D Solid & Generate building massing & Utilizing BuildingNet dataset~\cite{selvaraju2021buildingnet} as a basis to train urban morphology data to generate 3D building massing & - & GAN\\ \hline

\cite{miguel_deep_2019} & 3D Solid& Generation, manipulation and form finding of structural typologies & Utilizing VAE to learn continuous latent space to generate new geometries & Voxel grid & VAE\\ \hline
    
\cite{kahraman_augmenting_2021} & 3D Solid & Solve design problems incorporating deep learning & Develop post-integration in the generation workflow using VAE and GAN to solve complex multi-objective with abstract design criteria & Voxel grid & Base on VAE \& GAN\\ \hline
    
\cite{sebestyen_towards_2021} & 3D Solid & New way to design models & Utilizing VAE to encode and decode geometries through dimensionality manipulation & Voxel grid & VAE\\ \hline

\cite{liu_diffusion_2024} & 3D Solid & Form finding to assist in design & Developing a tool with a user interface (UI) to generate 3D point clouds using a diffusion model. & Point cloud & Diffusion model\\ \hline

\cite{sebestyen2023gaining} & 3D Solid & Form finding to assist design & Utilizing diffusion model to generate 3D forms with voxel grids through parametric design as dataset& Voxel grid & Diffusion model\\ \toprule

\cite{guida_gaining_nodate} & Text-3D Solid & Multimodal generation with text input & Utilizing diffusion model and NeRF to generate 3D forms from text input with multimodality & Mesh & Diffusion model \& NeRF\\ \hline

\cite{koh_ai-bewitched_2023} & Text-3D Solid & Test the connection between text and 3D form & Employing a specialized prompt to manipulate DreamField to generate multiple design alternatives for architects & Mesh & Diffusion model\\ \hline


\cite{guida_multimodal_2023} & Text-3D Solid & Multimodal generation with text input & Utilizing diffusion model to generate 3D forms from text input with multimodality & - & Diffusion model\\ 

\bottomrule

  \end{tabular}
    \label{tab:arch}
\end{table}

\end{document}